# Influences of $\delta B$ Contribution and Parallel Inertial Term of Energetic Particles on MHD-Kinetic Hybrid Simulations: A Case Study of the 1/1 Internal Kink Mode

H.X. Zhang[1], H.W. Zhang[2], Z.W. Ma[1,*], C. Liu[3]

[1] Institute for Fusion Theory and Simulation, Zhejiang University, Hangzhou 310027, China

[2] Max Planck Institute for Plasma Physics, Boltzmannstr. 2, 85748 Garching b. M., Germany

[3] Princeton Plasma Physics Laboratory, Princeton, New Jersey 08540, USA

## Abstract

The magnetohydrodynamic-kinetic (MHD-kinetic) hybrid model [Park et. al., 1992] has been widely applied in studying energetic particles (EPs) problems in fusion plasmas for past decades. The pressure-coupling scheme or the current-coupling scheme is adopted in this model. However, two noteworthy issues arise in the model application: firstly, the coupled term introduced in the pressure-coupling scheme, $(\mathbf{\nabla}\cdot\mathbf{P}_{\mathrm{h}})_{\perp}$, is often simplified by $\mathbf{\nabla}\cdot\mathbf{P}_{\mathrm{h}}$, which is equivalent to neglecting the parallel inertial term of EPs; secondly, besides the $\delta f$ contribution caused by changing in the EP distribution function, the magnetic field perturbation (the $\delta\boldsymbol{B}$ contribution) generated during development of the instabilities should also be considered, but it is often ignored in existing hybrid simulations. In this paper, we derive the analytical formulations under these two coupling schemes and then numerically study the representative case of the linear stability of the $m/n = 1/1$ internal kink mode (IKM) [Fu et. al., 2006] by using the CLT-K code. It is found that the approximated models can still yield reasonable results when EPs are isotopically distributed. But it fails completely in cases with anisotropic EP distributions. In addition, we further investigate the influence of EP's orbit width on the stability of IKM and verify the equivalence between pressure-coupling scheme and the current-coupling scheme.

* Corresponding Author: zwma@zju.edu.cn

## 1. Introduction

In fusion plasma simulations, the magnetohydrodynamic-kinetic (MHD-kinetic) hybrid model proposed by Park et. al. in the 1990s [1] has been widely used. The model (1) uses MHD equations to evolve the background plasma, and (2) solves the kinetic equation, usually via particle-in-cell (PIC) method, to evolve the distribution $f(\boldsymbol{x}, \boldsymbol{v}, t)$ of energetic particles (EPs). Compared with pure MHD simulations, hybrid simulations can include interactions between EPs and background plasma, e.g., through (inverse) Landau damping [2,3]. Meanwhile, hybrid simulations can solve the evolution of the background plasma more efficiently compared to full PIC simulations. EPs are expected to play an important role in future tokamak burning plasmas [4,5]. Over the past few decades, several hybrid codes, such as HGMC [6], MEGA [7], M3D-K [8,9], NIMORD [10,11], CLT-K [12], M3D-C1-K [13], JOREK [14], etc., have been developed to investigate EP-driven Alfvén eigenmodes (AEs) [12,15,16], energetic particle modes (EPMs) [12,17] , the influence of EPs on MHD instabilities [16,18], etc.

There are two main schemes used in hybrid simulations: the pressure-coupling scheme and the current-coupling scheme. In these two schemes, the contributions of EPs are coupled to the momentum equation of the background plasma in the form of $(\boldsymbol{\nabla} \cdot \mathbf{P}_\mathrm{h})_\perp$ or $\boldsymbol{J}_\mathrm{h} \times \boldsymbol{B}$, where $\mathbf{P}_\mathrm{h}$ and $\boldsymbol{J}_\mathrm{h}$ represent respectively the pressure tensor and current contributed by EPs. Analytically, these two schemes are strictly equivalent [18]. However, in numerical applications of the hybrid model, there are two issues worthy of attention:

First, instead of the standard pressure coupling term $(\boldsymbol{\nabla} \cdot \mathbf{P}_\mathrm{h})_\perp$, many pressure-coupling codes [9,11,13] adopt a simplified form as $\boldsymbol{\nabla} \cdot \mathbf{P}_\mathrm{h}$. This $(\boldsymbol{\nabla} \cdot \mathbf{P}_\mathrm{h})$-coupling scheme introduces a difference of $(\boldsymbol{\nabla} \cdot \mathbf{P}_\mathrm{h})_\parallel$, which equivalently neglects the parallel inertial term of EPs. The corresponding simulation results may differ from that obtained from standard pressure coupling using $(\boldsymbol{\nabla} \cdot \mathbf{P}_\mathrm{h})_\perp$, and will naturally be inconsistent with the current-coupling scheme. This issue was also highlighted in Park's original paper [1]. Thus, a quantitative comparison of these two different pressure-coupling schemes is necessary.

The second issue, independent of pressure-coupling or current-coupling schemes, only exists in the hybrid codes using the $\delta f$ method [19-21], which aims to reduce PIC noise by evolving only the perturbation of the distribution function. In such cases, the integrated coupling term is divided into two parts [18], i.e., the change of EPs' distribution function (the $\delta f$ contribution) and the perturbations of the magnetic field (the $\delta\boldsymbol{B}$ contribution) generated by instabilities. In many hybrid simulations [11,13], the influence of the $\delta\boldsymbol{B}$ contribution is also ignored, although it could be significant in certain situations.

This paper focuses on the above two issues, that is, whether these approximations in coupling schemes significantly influence the hybrid simulation results. We used the latest version of the CLT-K code [18], which supports both pressure-coupling and current-coupling schemes. We chose influence of EPs on the $m/n = 1/1$ internal kink mode [9] (IKM) as the subject (where $m$ is the poloidal mode number and $n$ is the toroidal one), and analyzed the differences in linear properties of IKM caused by different treatments in EP parallel inertial term and $\delta\boldsymbol{B}$ contribution. Some of the results were cross checked with the M3D-C1-K code. We further analyzed the reasons why previous simulations with simplified coupling schemes could still achieve results that are relatively consistent with the correct coupling schemes. Furthermore, we proved both theoretically and numerically that the pressure-coupling scheme and current-coupling scheme are strictly equivalent, regardless of whether the $\delta\boldsymbol{B}$ contribution terms are included.

The structure of this paper is as follows: Section 2 analyzes the MHD-kinetic hybrid model, highlighting differences in coupling schemes, such as the $\delta\boldsymbol{B}$ contribution and EP parallel inertial term, and gives comparison of the coupling schemes with/without simplification. Section 3 briefly overviews the models and numerical methods in CLT-K. Section 4 presents simulation results of the influence of nearly isotropic EPs on IKMs under different coupling schemes. The cases with anisotropic EPs will be discussed in Section 5. A preliminary investigation into the influence of EP Larmor radius will be conducted in Section 6. Finally, Section 7 summarizes and discusses the entire paper.

## 2. MHD-Kinetic Hybrid Model

### 2.1 Origins: Pressure-coupling and Current-coupling

In the MHD-kinetic hybrid model initially proposed by Park et al. [1], all contributions of EPs are attributed to the pressure tensor $\mathbf{P}_{\mathrm{h}}$ or current $\boldsymbol{J}_{\mathrm{h}}$ of EPs and coupled into the momentum equation of background plasma. As shown in Equations (1) or (2). The subscript 'h' represents EPs.

$$\rho\frac{\mathrm{d}\boldsymbol{V}}{\mathrm{d}t} = \boldsymbol{J}\times\boldsymbol{B} - \nabla p - (\nabla\cdot\mathbf{P}_{\mathrm{h}})_{\perp} \tag{1}$$

$$\rho\frac{\mathrm{d}\boldsymbol{V}}{\mathrm{d}t} = (\boldsymbol{J}-\boldsymbol{J}_{\mathrm{h}})\times\boldsymbol{B} - \nabla p \tag{2}$$

where $\frac{\mathrm{d}}{\mathrm{d}t} = \frac{\partial}{\partial t} + \boldsymbol{V}\cdot\nabla$ means the full derivative, $\rho$, $p$, $\boldsymbol{V}$, $\boldsymbol{B}$, $\boldsymbol{E}$ are plasma density, plasma pressure, fluid velocity, magnetic field, and electric field, respectively, and $\boldsymbol{J} = \boldsymbol{J}_{\mathrm{b}} + \boldsymbol{J}_{\mathrm{h}}$ represents the total plasma current (where $\boldsymbol{J}_\mathrm{b}$ represents the background plasma current). The above two schemes are called the "pressure-coupling scheme" and the "current-coupling scheme", respectively. Equation (1) is derived by subtracting the parallel component of the momentum equation of EPs from the momentum equation of the whole plasma, while Equation (2) is obtained by subtracting the complete momentum equation of EPs from the momentum equation of the whole plasma. In Park's derivation of the pressure-coupling scheme, the momentum evolution of the EPs in the perpendicular direction is neglected, that is, $\partial(\rho_\mathrm{h}\boldsymbol{V}_{\mathrm{h}\perp})/\partial t = \mathbf{0}$. In addition, here in the current-coupling scheme, the electric field force term $-q_\mathrm{h}\boldsymbol{E}_\perp(= q_\mathrm{h}\boldsymbol{V}\times\boldsymbol{B})$ (where $q_\mathrm{h}$ is the charge of the EP) does not appear in Equation (2), which is canceled by the current contributed by $\boldsymbol{E}\times\boldsymbol{B}$ drift of EPs. All instances of "perpendicular" and "parallel" are defined relative to the direction of the magnetic field $\boldsymbol{B}$.

Next, the update and calculation of $\mathbf{P}_\mathrm{h}$ or $\boldsymbol{J}_\mathrm{h}$ involve PIC simulations to evolve the EP distribution function $f$ and integrate it in the velocity space. The mass, velocity, and magnetic moment of EPs are denoted as $m_\mathrm{h}$, $\boldsymbol{v}$, and $\mu$, respectively, where $\mu = m_\mathrm{h}v_\perp^2/2B$.

In the pressure-coupling scheme, we write $\mathbf{P}_\mathrm{h}$ in Chew-Goldberger-Low (CGL) form [22]:

$$\mathbf{P}_\mathrm{h} = P_{\mathrm{h}\perp}\mathbf{I} + (P_{\mathrm{h}\parallel} - P_{\mathrm{h}\perp})\boldsymbol{bb} \tag{3}$$

where $\mathbf{I}$ is the unit tensor and $\boldsymbol{b} = \boldsymbol{B}/B$ is the unit vector along the direction of the magnetic field.

In the current-coupling scheme, the EP current $\boldsymbol{J}_\mathrm{h}$ mainly consists of two parts: the guiding center current and the magnetization current. Specifically, the former is contributed by the EP guiding center velocity (without $\boldsymbol{E}\times\boldsymbol{B}$ drift as explained above, and polarization drift is neglected in this work), including EP field-aligned velocity $\boldsymbol{v}_{\boldsymbol{b}} = v_\parallel\boldsymbol{b}$, magnetic field gradient drift velocity $\boldsymbol{v}_{\nabla B} = (\mu/q_\mathrm{h}B)\boldsymbol{b}\times\nabla B$ and magnetic field curvature drift velocity $\boldsymbol{v}_{\nabla\times\boldsymbol{b}} = \left(m_\mathrm{h}v_\parallel^2/q_\mathrm{h}B\right)\nabla\times\boldsymbol{b}$. The magnetization current is determined by the total magnetization $\boldsymbol{M}_\mathrm{h} = -\int\mu\boldsymbol{b}f\mathrm{d}\boldsymbol{v}$ of the EPs. Then the current of EPs in Equation (2) can be expressed as: [23]

$$\begin{aligned}\boldsymbol{J}_\mathrm{h} &= \int q_\mathrm{h}\left(v_\parallel\boldsymbol{b} + \frac{m_\mathrm{h}v_\parallel^2}{q_\mathrm{h}B}\nabla\times\boldsymbol{b} + \frac{\mu}{q_\mathrm{h}B}\boldsymbol{b}\times\nabla B\right)f\mathrm{d}\boldsymbol{v} + \nabla\times\left(-\int\mu\boldsymbol{b}f\mathrm{d}\boldsymbol{v}\right)\\ &= q_\mathrm{h}\widetilde{NV}_{\mathrm{h}\parallel}\boldsymbol{b} + \frac{1}{B}(P_{\mathrm{h}\parallel} - P_{\mathrm{h}\perp})\nabla\times\boldsymbol{b} + \frac{1}{B}\boldsymbol{b}\times\nabla P_{\mathrm{h}\perp}\end{aligned} \tag{4}$$

In Equations (3) and (4),

$$P_{\mathrm{h}\parallel} = \int m_\mathrm{h}v_\parallel^2 f\mathrm{d}\boldsymbol{v} \tag{5}$$

$$P_{\mathrm{h}\perp} = \int \mu B f\mathrm{d}\boldsymbol{v} \tag{6}$$

$$\widetilde{NV}_{\mathrm{h}\parallel} = \int v_{\parallel} f \mathrm{d}\boldsymbol{v} \tag{7}$$

## 2.2 Equivalence of the Two Coupling

The equivalence of the pressure-coupling scheme and the current-coupling scheme can be analytically proven as follows:

$$\begin{aligned}\boldsymbol{J}_{\mathrm{h}} \times \boldsymbol{B} &= \left[q_{\mathrm{h}} \widetilde{NV}_{\mathrm{h}\parallel} B \boldsymbol{b} + (P_{\mathrm{h}\parallel} - P_{\mathrm{h}\perp}) \nabla \times \boldsymbol{b} + \boldsymbol{b} \times \nabla P_{\mathrm{h}\perp}\right] \times \boldsymbol{b} \\ &= (\nabla P_{\mathrm{h}\perp})_{\perp} + (P_{\mathrm{h}\parallel} - P_{\mathrm{h}\perp})(\boldsymbol{b} \cdot \nabla)\boldsymbol{b}\end{aligned} \tag{8}$$

Simultaneously,

$$\begin{aligned}\nabla \cdot \mathbf{P}_{\mathrm{h}} &= \nabla \cdot [P_{\mathrm{h}\perp}\mathbf{I} + (P_{\mathrm{h}\parallel} - P_{\mathrm{h}\perp})\boldsymbol{bb}] \\ &= (\nabla P_{\mathrm{h}\perp})_{\perp} + (\nabla P_{\mathrm{h}\perp})_{\parallel} + (P_{\mathrm{h}\parallel} - P_{\mathrm{h}\perp})(\boldsymbol{b} \cdot \nabla)\boldsymbol{b} \\ &\qquad + [(P_{\mathrm{h}\parallel} - P_{\mathrm{h}\perp})(\nabla \cdot \boldsymbol{b}) + \boldsymbol{b} \cdot \nabla(P_{\mathrm{h}\parallel} - P_{\mathrm{h}\perp})]\boldsymbol{b}\end{aligned} \tag{9}$$

$$(\nabla \cdot \mathbf{P}_{\mathrm{h}})_{\perp} = (\nabla P_{\mathrm{h}\perp})_{\perp} + (P_{\mathrm{h}\parallel} - P_{\mathrm{h}\perp})(\boldsymbol{b} \cdot \nabla)\boldsymbol{b} \tag{10}$$

Therefore, it can be concluded from Equations (8) ~ (10) that the EP contribution terms subtracted from the right side of the total momentum equation are equal in two coupling schemes: [18]

$$\boldsymbol{J}_{\mathrm{h}} \times \boldsymbol{B} = (\nabla \cdot \mathbf{P}_{\mathrm{h}})_{\perp} \tag{11}$$

That is, Equation (1) is equivalent to Equation (2). Clearly, neither the parallel components of $\boldsymbol{J}_{\mathrm{h}}$ nor $\nabla \cdot \mathbf{P}_{\mathrm{h}}$ contribute to the momentum equation.

## 2.3 Perturbation Forms of the Momentum Equation

In PIC simulations, the total distribution function $f$ of EPs can be divided into initial and perturbed components, that is: $f = f_0 + \delta f$. In the $\delta f$ method [19,20] adopted by CLT-K and many other codes [7,9,11-13], only the $\delta f$ components evolve. Similarly, all quantities on grids ($\rho$, $p$, $\boldsymbol{V}$, $\boldsymbol{B}$, $\boldsymbol{E}$, $\boldsymbol{J}$, $\mathbf{P}_{\mathrm{h}}$, $\boldsymbol{J}_{\mathrm{h}}$) can also be decomposed into equilibrium (initial) and perturbation components, e.g., $\boldsymbol{B} = \boldsymbol{B}_0 + \delta\boldsymbol{B}$, where only the perturbation components evolve. This approach can reduce numerical errors caused by limited accuracy in the initial equilibrium, which is pre-determined by equilibrium codes [24-26] by solving the Grad-Shafranov equation.

For hybrid simulations, in principle, the EP contribution to the initial equilibrium should be included. If the initial equilibrium is static (i.e., $\boldsymbol{V}_0 = \mathbf{0}$), then the total plasma satisfies:

$$\boldsymbol{J}_0 \times \boldsymbol{B}_0 - \nabla p_0 - (\nabla \cdot \mathbf{P}_{\mathrm{h}0})_{\perp 0} = \mathbf{0} \tag{12}$$

where the subscript '$\perp 0$' means perpendicular to $\boldsymbol{b}_0$. Therefore, subtracting Equation (12) from Equation (1) yields the perturbed form of the momentum equation in the pressure-coupling scheme:

$$\rho\frac{\mathrm{d}\boldsymbol{V}}{\mathrm{d}t}=\boldsymbol{J}_0\times\delta\boldsymbol{B}+\delta\boldsymbol{J}\times\boldsymbol{B}-\nabla\delta p-\delta[(\nabla\cdot\mathbf{P}_\mathrm{h})_\perp] \tag{13}$$

where

$$\begin{aligned}\delta[(\nabla\cdot\mathbf{P}_\mathrm{h})_\perp]&=(\nabla\cdot\mathbf{P}_\mathrm{h})_\perp-(\nabla\cdot\mathbf{P}_\mathrm{h0})_{\perp0}\\&=(\nabla\cdot\delta\mathbf{P}_\mathrm{h})_\perp+(\nabla\cdot\mathbf{P}_\mathrm{h0})_\perp-(\nabla\cdot\mathbf{P}_\mathrm{h0})_{\perp0}\end{aligned} \tag{14}$$

and

$$\begin{aligned}\delta\mathbf{P}_\mathrm{h}&=\mathbf{P}_\mathrm{h}-\mathbf{P}_\mathrm{h0}\\&=[P_{\mathrm{h}\perp}\mathbf{I}+(P_{\mathrm{h}\parallel}-P_{\mathrm{h}\perp})\boldsymbol{bb}]-[P_{\mathrm{h}\perp0}\mathbf{I}+(P_{\mathrm{h}\parallel0}-P_{\mathrm{h}\perp0})\boldsymbol{b}_0\boldsymbol{b}_0]\\&=\delta P_{\mathrm{h}\perp}\mathbf{I}+(\delta P_{\mathrm{h}\parallel}-\delta P_{\mathrm{h}\perp})\boldsymbol{bb}+(P_{\mathrm{h}\parallel0}-P_{\mathrm{h}\perp0})(\boldsymbol{b}_0\delta\boldsymbol{b}+\delta\boldsymbol{b}\boldsymbol{b})\end{aligned} \tag{15}$$

Equivalently, the initial equilibrium of the background plasma should satisfy:

$$(\boldsymbol{J}_0-\boldsymbol{J}_\mathrm{h0})\times\boldsymbol{B}_0-\nabla p_0=\mathbf{0} \tag{16}$$

Subtracting Equation (16) from Equation (2) leads to the perturbation form of the momentum equation in the current-coupling scheme:

$$\rho\frac{\mathrm{d}\boldsymbol{V}}{\mathrm{d}t}=\boldsymbol{J}_0\times\delta\boldsymbol{B}-\boldsymbol{J}_\mathrm{h0}\times\delta\boldsymbol{B}+(\delta\boldsymbol{J}-\delta\boldsymbol{J}_\mathrm{h})\times\boldsymbol{B}-\nabla\delta p \tag{17}$$

where

$$\begin{aligned}\delta\boldsymbol{J}_\mathrm{h}&=\boldsymbol{J}_\mathrm{h}-\boldsymbol{J}_\mathrm{h0}\\&=q_\mathrm{h}\delta\big(\widetilde{N}V_{\mathrm{h}\parallel}\big)\boldsymbol{b}+\frac{1}{B}(\delta P_{\mathrm{h}\parallel}-\delta P_{\mathrm{h}\perp})\nabla\times\boldsymbol{b}+\frac{1}{B}\boldsymbol{b}\times\nabla\delta P_{\mathrm{h}\perp}\\&\quad+q_\mathrm{h}\big(\widetilde{N}V_{\mathrm{h}\parallel}\big)_0\delta\boldsymbol{b}+\frac{1}{BB_0}(P_{\mathrm{h}\parallel0}-P_{\mathrm{h}\perp0})(B_0\nabla\times\delta\boldsymbol{b}-\delta B\nabla\times\boldsymbol{b}_0)\\&\quad+\frac{1}{BB_0}(B_0\delta\boldsymbol{b}-\delta B\boldsymbol{b}_0)\times\nabla P_{\mathrm{h}\perp0}\end{aligned} \tag{18}$$

In these equations, $\mathbf{P}_\mathrm{h0}$, $\boldsymbol{J}_\mathrm{h0}$, $P_{\mathrm{h}\parallel0}$, $P_{\mathrm{h}\perp0}$, $\big(\widetilde{N}V_{\mathrm{h}\parallel}\big)_0$ and $\delta P_{\mathrm{h}\parallel}$, $\delta P_{\mathrm{h}\perp}$, $\delta\big(\widetilde{N}V_{\mathrm{h}\parallel}\big)$ can be respectively obtained by integrating over $f_0$ and $\delta f$, similar to Equations (3) ~ (7).

Most equilibrium codes use isotropic scalar pressure for simplicity [24-26]. However, the anisotropic EP distribution, common in strongly magnetized plasmas, makes it challenging to solve the self-consistent equilibria including EP contributions. Thus, the equilibria based on isotropic scalar pressure lead to the mismatch in Equations (12) and (16), which is artificially ignored by adopting the perturbation forms of the momentum equations, i.e., Equations (13) and (17). [a] Self-consistently solving the initial equilibrium in with anisotropic EPs is beyond the scope of this paper.

[a] To avoid this situation, the contribution of EPs to the equilibrium has been completely ignored in many past simulations, but this will also introduce new errors. In specific simulations, by carefully using an isotropic EP distribution function with the same radial profile as the background plasma can simplify EP equilibrium pressure $\mathbf{P}_\mathrm{h0}$ to an isotropic scalar pressure $p_\mathrm{h0}$, ensuring strict initial mechanical equilibrium [11] . Still, this approach may not be suitable for more general cases.

## 2.4 Issue 1 : the $\delta B$ Contribution

In Equations (13) ~ (15) and (17), (18), it is evident that contributions related to EPs can be divided into two parts [18]: One originates from $\delta f$, the perturbation of the EP distribution function, which is coupled to the momentum equation through terms such as $\delta P_{\mathrm{h}\parallel}$, $\delta P_{\mathrm{h}\perp}$, and $\delta(\widetilde{NV}_{\mathrm{h}\parallel})$, and we refer to this contribution as the "$\delta f$ contribution", represented by unhighlighted terms in Equations (14) ~ (15) and (18). The other part comes from the contribution of the EP initial distribution function $f_0$ (manifesting as $\boldsymbol{J}_{\mathrm{h}_0}$ or $\mathbf{P}_{\mathrm{h}_0}$), to the magnetic field perturbation $\delta\boldsymbol{B}$. We call this contribution the "$\delta\boldsymbol{B}$ contribution", which is represented by terms in Equations (14), (15) and (17), (18) highlighted in blue.

Both $\delta f$ and $\delta\boldsymbol{B}$ contribution terms are first-order. Therefore, for the self-consistency and rigor of the hybrid model, in principle, both of these two effects should be considered in the momentum equation, regardless of coupling scheme chosen.

### 2.4.1 Model without $\delta B$ Terms

Nevertheless, in most simulations of EP-induced AEs and EPMs, it is often believed that the $\delta f$ contribution caused by non-adiabatic wave-particle interaction dominates over the disturbance of the magnetic field (especially in the linear phase). So, the $\delta\boldsymbol{B}$ contribution is frequently overlooked, and only the $\delta f$ contribution is considered, leading to further simplification of the model.

For the case without the $\delta\boldsymbol{B}$ contribution, the momentum equation (13) in the pressure-coupling scheme will be simplified to [for distinction, we add a superscript '$(\delta f)$']:

$$\rho\frac{\mathrm{d}\boldsymbol{V}}{\mathrm{d}t}=\boldsymbol{J}_0\times\delta\boldsymbol{B}+\delta\boldsymbol{J}\times\boldsymbol{B}-\nabla\delta p-(\nabla\cdot\delta\mathbf{P}_{\mathrm{h}})_{\perp}^{(\delta f)} \tag{19}$$

where $(\nabla\cdot\delta\mathbf{P}_{\mathrm{h}})_{\perp}^{(\delta f)}$ is used instead of $\delta[(\nabla\cdot\mathbf{P}_{\mathrm{h}})_{\perp}]$, and

$$\delta\mathbf{P}_{\mathrm{h}}^{(\delta f)}=\delta P_{\mathrm{h}\perp}\mathbf{I}+(\delta P_{\mathrm{h}\parallel}-\delta P_{\mathrm{h}\perp})\boldsymbol{bb} \tag{20}$$

Similarly, the momentum equation (17) in the current-coupling scheme will be simplified to

$$\rho\frac{\mathrm{d}\boldsymbol{V}}{\mathrm{d}t}=\boldsymbol{J}_0\times\delta\boldsymbol{B}+\left(\delta\boldsymbol{J}-\delta\boldsymbol{J}_{\mathrm{h}}^{(\delta f)}\right)\times\boldsymbol{B}-\nabla\delta p \tag{21}$$

where

$$\delta\boldsymbol{J}_{\mathrm{h}}^{(\delta f)}=q_{\mathrm{h}}\delta(\widetilde{NV}_{\mathrm{h}\parallel})\boldsymbol{b}+\frac{1}{B}(\delta P_{\mathrm{h}\parallel}-\delta P_{\mathrm{h}\perp})\nabla\times\boldsymbol{b}+\frac{1}{B}\boldsymbol{b}\times\nabla\delta P_{\mathrm{h}\perp} \tag{22}$$

Similar to Equations (8) to (11), we can also analytically prove the strict equivalence of the pressure-coupling and the current-coupling schemes in Equations (19) and (21), i.e.,

$$\delta \boldsymbol{J}_{\mathrm{h}}^{(\delta f)} \times \boldsymbol{B} = (\boldsymbol{\nabla} \cdot \delta \mathbf{P}_{\mathrm{h}})_{\perp}^{(\delta f)}$$

$$= (\boldsymbol{\nabla} \delta P_{\mathrm{h}\perp})_{\perp} + (\delta P_{\mathrm{h}\parallel} - \delta P_{\mathrm{h}\perp})(\boldsymbol{b} \cdot \boldsymbol{\nabla})\boldsymbol{b} \tag{23}$$

This equivalence has also been further verified with numerical simulation results of a toroidal Alfvén eigenmode (TAE), as shown in Figure 1 of the Reference [18].

#### 2.4.2 Model with $\delta\boldsymbol{B}$ Terms

In addition to the $\delta f$ contribution, the correction brought by the magnetic field perturbation, $\delta\boldsymbol{B}$, can also significantly affect the simulation results in cases where the magnetic surface structure exhibits strong deformation, such as of low-frequency MHD instabilities like IKMs and tearing modes (TMs), or even in the nonlinear phase of AEs. A more complete model would need to take into account the $\delta\boldsymbol{B}$ contribution.

For the case with the $\delta\boldsymbol{B}$ contribution, the equivalence between the current-coupling scheme and the pressure-coupling scheme still holds. Regarding this, Reference [18] used a TAE simulation result to verify (see Figure 2 of it), which also shows the $\delta\boldsymbol{B}$ contribution will bring significant differences to the simulation results of the TAE nonlinear phase. However, no detailed analytical proof is given. In Appendix A of this paper, we provide an analytical proof of this equivalence: the additional terms contributed by EPs in both schemes are equal, namely

$$\delta[(\boldsymbol{\nabla} \cdot \mathbf{P}_{\mathrm{h}})_{\perp}] = \delta \boldsymbol{J}_{\mathrm{h}} \times \boldsymbol{B} + \boldsymbol{J}_{\mathrm{h}0} \times \delta \boldsymbol{B}$$

$$= (\boldsymbol{\nabla} \delta P_{\mathrm{h}\perp})_{\perp} + (\delta P_{\mathrm{h}\parallel} - \delta P_{\mathrm{h}\perp})(\boldsymbol{b} \cdot \boldsymbol{\nabla})\boldsymbol{b}$$

$$+(\boldsymbol{\nabla} P_{\mathrm{h}\perp 0})_{\perp} - (\boldsymbol{\nabla} P_{\mathrm{h}\perp 0})_{\perp 0} + (P_{\mathrm{h}\parallel 0} - P_{\mathrm{h}\perp 0})[(\boldsymbol{b} \cdot \boldsymbol{\nabla})\boldsymbol{b} - (\boldsymbol{b}_0 \cdot \boldsymbol{\nabla})\boldsymbol{b}_0] \tag{24}$$

#### 2.4.3 Essence of the $\delta f$ and $\delta\boldsymbol{B}$ Contributions

Starting from now, we will take the pressure-coupling scheme as our starting point for analysis. According to the results of Equation (24), in the complete coupling model, the contributions $\delta[(\boldsymbol{\nabla} \cdot \mathbf{P}_{\mathrm{h}})_{\perp}]$ from EPs include three parts:

$$\delta[(\boldsymbol{\nabla} \cdot \mathbf{P}_{\mathrm{h}})_{\perp}] = (\boldsymbol{\nabla} \cdot \delta \mathbf{P}_{\mathrm{h}})_{\perp}^{(\delta f)} + \delta[(\boldsymbol{\nabla} \cdot \mathbf{P}_{\mathrm{h}})_{\perp}]^{(\delta \boldsymbol{B},\mathrm{perp})}$$

$$+ \delta[(\boldsymbol{\nabla} \cdot \mathbf{P}_{\mathrm{h}})_{\perp}]^{(\delta \boldsymbol{B},\mathrm{cur})} \tag{25}$$

in which,

$$(\boldsymbol{\nabla} \cdot \delta \mathbf{P}_{\mathrm{h}})_{\perp}^{(\delta f)} = (\boldsymbol{\nabla} \delta P_{\mathrm{h}\perp})_{\perp} + (\delta P_{\mathrm{h}\parallel} - \delta P_{\mathrm{h}\perp})\boldsymbol{\kappa} \tag{26}$$

$$\delta[(\boldsymbol{\nabla} \cdot \mathbf{P}_{\mathrm{h}})_{\perp}]^{(\delta \boldsymbol{B},\mathrm{perp})} = (\boldsymbol{\nabla} P_{\mathrm{h}\perp 0})_{\perp} - (\boldsymbol{\nabla} P_{\mathrm{h}\perp 0})_{\perp 0} = \delta[(\boldsymbol{\nabla} P_{\mathrm{h}\perp 0})_{\perp}] \tag{27}$$

$$\delta[(\boldsymbol{\nabla} \cdot \mathbf{P}_{\mathrm{h}})_{\perp}]^{(\delta \boldsymbol{B},\mathrm{cur})} = (P_{\mathrm{h}\parallel 0} - P_{\mathrm{h}\perp 0})(\boldsymbol{\kappa} - \boldsymbol{\kappa}_0) = (P_{\mathrm{h}\parallel 0} - P_{\mathrm{h}\perp 0})\delta \boldsymbol{\kappa} \tag{28}$$

where $\boldsymbol{\kappa} = (\boldsymbol{b} \cdot \boldsymbol{\nabla})\boldsymbol{b}$ represents the curvature of the magnetic field.

Equations (25) mean that the $\delta \boldsymbol{B}$ contribution of EPs can be intuitively divided into two parts, $\delta[(\nabla \cdot \mathbf{P}_\mathrm{h})_\perp]^{(\delta \boldsymbol{B},\mathrm{perp})}$ ("perp-terms") and $\delta[(\nabla \cdot \mathbf{P}_\mathrm{h})_\perp]^{(\delta \boldsymbol{B},\mathrm{cur})}$ ("cur-terms"), comes from the change in the direction and local curvature of the magnetic field, respectively. Additionally, it can be seen clearly that the $\delta \boldsymbol{B}$ contribution is of the same order as the $\delta f$ contribution, i.e., $(\nabla \cdot \delta \mathbf{P}_\mathrm{h})_\perp^{(\delta f)}$. Since $\delta[(\nabla \cdot \mathbf{P}_\mathrm{h})_\perp]^{(\delta \boldsymbol{B},\mathrm{cur})}$ is proportional to $(P_{\mathrm{h}\parallel 0} - P_{\mathrm{h}\perp 0})$, it strongly depends on the anisotropy of the initial EP equilibrium pressure. For cases where the initial EP equilibrium pressure is close to isotropic, the effect of this term will be negligible.

Generally, in most hybrid simulations based on the $\delta f$ method, the focus has primarily been on the $\delta f$ contribution, with less consideration given to the $\delta \boldsymbol{B}$ contribution. Only a few simulations based on the current-coupling scheme [7,13] consider the $\boldsymbol{J}_{\mathrm{h}0} \times \delta \boldsymbol{B}$ term, which is a partial $\delta \boldsymbol{B}$ contribution (often ignoring $\delta B$, the change of the magnetic field magnitude [23]). The importance of the $\delta \boldsymbol{B}$ contribution and its influence on simulation results is undoubtedly a matter of concern.

## 2.5 Issue 2 : Replacing $(\nabla \cdot \mathbf{P}_\mathrm{h})_\perp$ with $(\nabla \cdot \mathbf{P}_\mathrm{h})$

Some pressure-coupling schemes used in research [9,11,13] tend to subtract the entire $(\nabla \cdot \mathbf{P}_\mathrm{h})$ in the momentum equation for simplicity, rather than the perpendicular component $(\nabla \cdot \mathbf{P}_\mathrm{h})_\perp$. In this scenario, the momentum equation (1) becomes

$$\rho \frac{\mathrm{d}\boldsymbol{V}}{\mathrm{d}t} = \boldsymbol{J} \times \boldsymbol{B} - \nabla p - \nabla \cdot \mathbf{P}_\mathrm{h} \tag{29}$$

and its perturbation form is (of course, with only $\delta f$ terms)

$$\rho \frac{\mathrm{d}\boldsymbol{V}}{\mathrm{d}t} = \boldsymbol{J}_0 \times \delta \boldsymbol{B} + \delta \boldsymbol{J} \times \boldsymbol{B} - \nabla \delta p - \nabla \cdot \delta \mathbf{P}_\mathrm{h}^{(\delta f)} \tag{30}$$

The specific form of $\nabla \cdot \delta \mathbf{P}_\mathrm{h}^{(\delta f)}$ is

$$\begin{aligned} \nabla \cdot \delta \mathbf{P}_\mathrm{h}^{(\delta f)} = (\nabla \delta P_{\mathrm{h}\perp})_\perp &+ (\nabla \delta P_{\mathrm{h}\perp})_\parallel + (\delta P_{\mathrm{h}\parallel} - \delta P_{\mathrm{h}\perp})(\boldsymbol{b} \cdot \nabla)\boldsymbol{b} \\ &+ [(\delta P_{\mathrm{h}\parallel} - \delta P_{\mathrm{h}\perp})(\nabla \cdot \boldsymbol{b}) + \boldsymbol{b} \cdot \nabla(\delta P_{\mathrm{h}\parallel} - \delta P_{\mathrm{h}\perp})]\boldsymbol{b} \end{aligned} \tag{31}$$

The difference between it and $(\nabla \cdot \delta \mathbf{P}_\mathrm{h})_\perp^{(\delta f)}$ is the gray terms (also denoted as $(\nabla \cdot \delta \mathbf{P}_\mathrm{h})_\parallel^{(\delta f)}$). We call this coupling scheme the "$(\nabla \cdot \mathbf{P}_\mathrm{h})$-coupling scheme", distinguishing it from the "$(\nabla \cdot \mathbf{P}_\mathrm{h})_\perp$-coupling scheme", which is the standard pressure-coupling scheme originally proposed by Park et al. [1].

The $(\nabla \cdot \mathbf{P}_\mathrm{h})$-coupling scheme can be considered as an approximation that additionally neglects EP parallel inertial terms, i.e., $\partial(\rho_h \boldsymbol{V}_{h\parallel})/\partial t = \mathbf{0}$ [b]. It is also mentioned in Reference [1]: For the case when trapped particles dominate, this approximation may be feasible; but for passing particles, since the parallel inertia term of EP is usually not small, this approximation could lead to significant discrepancies and cannot be consistent with the current-coupling scheme. We believe that adopting the $(\nabla \cdot \mathbf{P}_h)_\perp$-coupling scheme is likely to derive more accurate results, but the specific differences between them still need to be thoroughly investigated.

### 2.6 Three Coupling Schemes Studied in this Paper

In this paper, we aim to comprehensively explore these two issues, i.e., the influence of the $\delta\boldsymbol{B}$ contribution and EP parallel inertia terms on MHD-kinetic simulations. In practice, we will use and compare the following three pressure coupling schemes (the corresponding current-coupling schemes can give equivalent results as the pressure coupling scheme A and B):

**[Scheme A]** Using the $(\nabla \cdot \mathbf{P}_h)_\perp$-coupling (or equivalent current-coupling) scheme with only the $\delta f$ contribution. In this case, the EP contribution is represented as $(\nabla \cdot \delta\mathbf{P}_h)_\perp^{(\delta f)}$.

**[Scheme B]** Using the standard $(\nabla \cdot \mathbf{P}_h)_\perp$-coupling (or equivalent current-coupling) scheme with both the $\delta f$ and $\delta\boldsymbol{B}$ contributions, which should be more accurate in principle. In this case, the EP contribution is expressed as $(\nabla \cdot \delta\mathbf{P}_h)_\perp^{(\delta f)} + \delta[(\nabla \cdot \mathbf{P}_h)_\perp]^{(\delta\boldsymbol{B},\text{perp})} + \delta[(\nabla \cdot \mathbf{P}_h)_\perp]^{(\delta\boldsymbol{B},\text{cur})}$.

**[Scheme C]** Using the $(\nabla \cdot \mathbf{P}_h)$-coupling scheme with only $\delta f$ contribution, which has been widely adopted in previous simulations based on the pressure-coupling scheme and has yielded generally reliable results. In this case, the EP contribution is described as $(\nabla \cdot \delta\mathbf{P}_h)_\perp^{(\delta f)} + (\nabla \cdot \delta\mathbf{P}_h)_\parallel^{(\delta f)}$.

All schemes can be readily implemented in the CLT-K code, and we will briefly introduce the code in Section 3. We choose a well-studied $m/n = 1/1$ IKM case [9] as our research subject, and the specific simulation results and analysis will be presented in Section 4 and Section 5.

## 3. Overview of the CLT-K Code

### 3.1 MHD Simulations in the CLT Code

CLT-K code was developed based on the 3D toroidal full MHD code CLT, which solves MHD

[b] For EPs, the parallel component of the momentum equation is $\partial(\rho_h \boldsymbol{V}_{h\perp})/\partial t = -(\nabla \cdot \mathbf{P}_h)_\parallel$. Therefore, ignoring the difference caused by term $(\nabla \cdot \mathbf{P}_h)_\parallel$ essentially amounts to neglecting the EPs' parallel inertial term $\partial(\rho_h \boldsymbol{V}_{h\perp})/\partial t$.

equations to evolve the background plasma in the cylindrical coordinate system $\{R, \phi, Z\}$ (where $R$ represents the major radius, $\phi$ is the toroidal angle, and $Z$ along the vertical direction). The equation sets and the dimensionless principles of various physical quantities ($\rho$, $p$, $\boldsymbol{V}$, $\boldsymbol{B}$, $\boldsymbol{E}$, $\boldsymbol{J}$, $D$, $\kappa$, $\nu$, $\eta$) are detailed in Reference [27], where $D$, $\kappa$, $\nu$, and $\eta$ are the dissipation coefficients (diffusivity, thermal conductivity, viscosity, and resistivity respectively). Notably, the Alfvén time is defined as $\tau_{\mathrm{A}} = a/v_{\mathrm{A}}$, where $a$ is tokamak's minor radius and $v_{\mathrm{A}}$ is the Alfvén speed at the magnetic axis. The definition of the Alfvén frequency is $\omega_{\mathrm{A}} = \tau_{\mathrm{A}}^{-1} = v_{\mathrm{A}}/a$.

In the currently used version of the CLT-K code, all directions are meshed uniformly. In addition, according to the specific needs of the physical problems, other modules can be enabled in the CLT code, such as the Hall effect [28], impurities, driven current [29], resonant magnetic perturbation (RMP) [30], etc. The code can achieve parallel acceleration on graphics processing units (GPUs) [31] and has been benchmarked against the M3D-C1 code [32].

## 3.2 MHD-Kinetic Hybrid Simulations in the CLT-K Code

Based on the CLT code, to include EP effects, the CLT-K code using the MHD-kinetic hybrid model was developed. [12,15,16,18]

The latest version of the CLT-K code supports both pressure-coupling and current-coupling schemes [16]. This involves coupling all EP contributions to the momentum equation of CLT in the form of the EP pressure tensor $\mathbf{P}_{\mathrm{h}}$ or EP current $\boldsymbol{J}_{\mathrm{h}}$, as summarized in Equation (1) or Equation (2) (with only an additional non-ideal viscous term $\boldsymbol{\nabla} \cdot [\nu \boldsymbol{\nabla}(\boldsymbol{V} - \boldsymbol{V}_0)]$), consistent with the original model in Reference [1]. The actual equations used in the code are still based on the $\delta f$ method and the perturbation form, i.e., Equation (13) or (17).

Here, we only briefly outline the three steps in CLT-K for obtaining $\boldsymbol{J}_{\mathrm{h}}$ or $\mathbf{P}_{\mathrm{h}}$. More detailed equations and elaborations can be found in Reference [18].

**Step 1: Pushing EPs in the electromagnetic field.** Based on the guiding-center motion equations [33], the electric field $\boldsymbol{E}$ and magnetic field $\boldsymbol{B}$ solved in the MHD equations drive EP motion in the four-dimensional phase space $\{\boldsymbol{X}, v_{\parallel}\}$, where $\boldsymbol{X}$ represents the guiding-center coordinates (when neglecting the Larmor radius, the magnetic moment $\mu$ is invariant).

**Step 2: Evolving the distribution function of EPs with $\boldsymbol{\delta f}$ method.** In CLT-K, PIC and $\delta f$ method [12,19,20] are used to evolve the distribution function of EPs. The latest version of CLT-K supports non-uniform sampling of markers, which increases the density of Monte Carlo samples in specific regions of the phase space, such as regions with strong wave-particle interactions or high

EP density. To achieve this, a function $g$ representing the density distribution of marker sampling particles is introduced [12,21,34], so that the weight of each marker's contribution to $\delta f$ is $w = \delta f / g$. Therefore, according to the Vlasov equation, that is, $\mathrm{d}f/\mathrm{d}t = 0$, the evolution of $w$ satisfies:

$$\frac{\mathrm{d}w}{\mathrm{d}t} = -\frac{1}{g}\frac{\mathrm{d}f_0}{\mathrm{d}t} = -\frac{1}{g}\left(\frac{\partial f_0}{\partial P_\phi}\frac{\mathrm{d}P_\phi}{\mathrm{d}t} + \frac{\partial f_0}{\partial \varepsilon}\frac{\mathrm{d}\varepsilon}{\mathrm{d}t} + \frac{\partial f_0}{\partial \Lambda}\frac{\mathrm{d}\Lambda}{\mathrm{d}t}\right) \tag{32}$$

where $P_\phi = m_\mathrm{h} v_\parallel R B_\phi / B - q_\mathrm{h}\psi$ is the poloidal canonical angular momentum ($\psi$ is the poloidal magnetic flux), $\varepsilon = \mu B + m_\mathrm{h} v_\parallel^2/2$ is the kinetic energy, and $\Lambda = \mu B_\mathrm{m}/\varepsilon$ is the pitch angle of EPs.

**Step 3: Integrating the distribution function of EPs to obtain the EP pressure or current.** The contributions from all markers are integrated (summed) to obtain $\mathbf{P}_\mathrm{h}$ or $\boldsymbol{J}_\mathrm{h}$, and then coupled to the momentum equation. Their complete forms and perturbation forms are the same as introduced in Section 2, represented respectively by Equations (3) ~ (7) and Equations (14) ~ (15) or (18).

# 4. Influences of Nearly Isotropic EPs on the Stability of the 1/1 IKM

## 4.1 Initial Equilibrium and Simulation Parameters

In this section, we will conduct simulations using the CLT-K code to revisit the well-known case studied by Fu et al. regarding the influence of nearly isotropically distributed EPs on the $m/n = 1/1$ IKM [9]. All parameters and initial equilibrium are consistent with Section 3 of Reference [9].

Specifically, we selected a circular tokamak geometry with an aspect ratio of $R_0/a = 2.763$. The total beta at the magnetic axis is $\beta_\mathrm{total0} = 8\%$ and the overall beta profile is:

$$\beta_\mathrm{total} = \beta_\mathrm{total0}\exp\left(-\frac{\bar{\psi}}{0.25}\right) \tag{33}$$

where $\bar{\psi} = \psi/(\psi_\mathrm{max} - \psi_\mathrm{min})$ refers to the normalized poloidal magnetic flux (with $\bar{\psi} = 0$ at the magnetic axis and $\bar{\psi} = 1$ at the boundary). The minor radius $r$ is approximated by $\sqrt{\bar{\psi}}$. This equilibrium is generated by NOVA's QSOLVER [24], as shown in Figure 1.

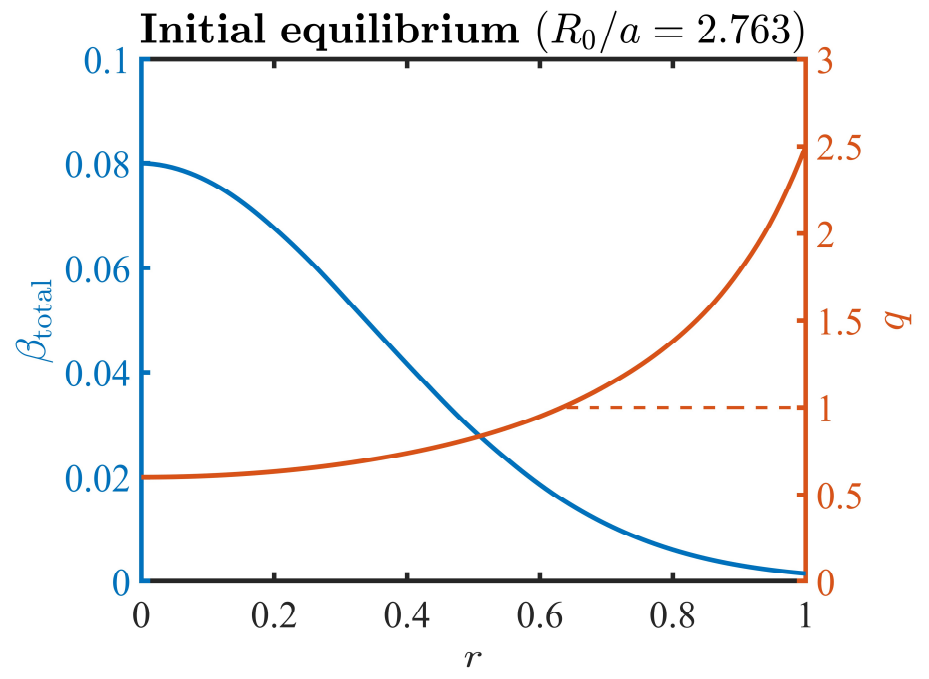

***Figure 1 :*** Profiles of $\beta_{\rm total}$ and the safety factor $q$ in the initial equilibrium, with a rational surface at $q = 1$ located at $r = 0.63$.

In the presence of EPs, we use the fraction of EP beta $\beta_{\rm h}$ to the total beta $\beta_{\rm total}$ to reflect the share of the EP pressure. The shapes of both the beta of EP ($\beta_{\rm h}$) and the beta of the background plasma ($\beta_{\rm b}$) match the total beta profile. The EPs initially satisfy a nearly isotropic [c] slowing-down distribution:

$$f_0 = \frac{1}{v^3 + v_{\rm c}^3}\left[1 + {\rm erf}\left(\frac{v_0 - v}{\Delta v}\right)\right]\exp\left(-\frac{\overline{\langle\psi\rangle}}{0.25}\right) \tag{34}$$

where the birth speed of EP is $v_0 = 4v_{\rm A}$, the critical speed is $v_{\rm c} = 0.58v_0$, $\Delta v = 0.001v_{\rm A}$, the Larmor radius of EP ($v_\perp = v_0$) is $\varrho_{\rm h} = 0.0125a$, and $\langle\psi\rangle = -P_\phi/q_{\rm h} + (m_{\rm h}/q_{\rm h})\langle v_\parallel RB_\phi/B\rangle$ denotes the average magnetic flux along the particle orbit (normalized as $\overline{\langle\psi\rangle}$). Note that the drift kinetic model without finite Larmor radius (FLR) effects is adopted, and the effect of different $\varrho_{\rm h}$ is mainly reflected in the orbit widths of EPs.

In the $\{R, \phi, Z\}$ directions, the grid resolution is $256 \times 16 \times 256$, totaling 4 million non-uniform sampled markers (the sampling weight $g$ in the phase space is proportional to the initial distribution function $f_0$). The convergence has been verified by scanning the grid resolution and the marker number.

## 4.2 Results of Pure MHD Simulations

First, for a pure MHD simulation without EP, an $m/n = 1/1$ IKM is destabilized inside the $q = 1$ rational surface, and its linear growth rate is about $\gamma = 0.0091$. The time evolution of the kinetic energy is shown by solid lines in Figure 2, the mode frequency is zero, and the mode structure is shown in Figure 3. The $m/n = 1/1$ radial displacement $|\xi_r|$ exhibits a step function, consistent with the characteristics of the IKM. In linear problems, the displacement $\xi$ is approximated as the integral of the fluid velocity, i.e., $\boldsymbol{\xi}(\boldsymbol{x},t)=\int_0^t \boldsymbol{v}(\boldsymbol{x},t')\mathrm{d}t'$.

[c] Here, "nearly isotropic" means that the EPs' distribution function does not explicitly depend on the pitch angle $\Lambda$. However, due to the definition of $\langle\psi\rangle$ involving $v_\parallel$, there are still minor differences from a strictly isotropic distribution.

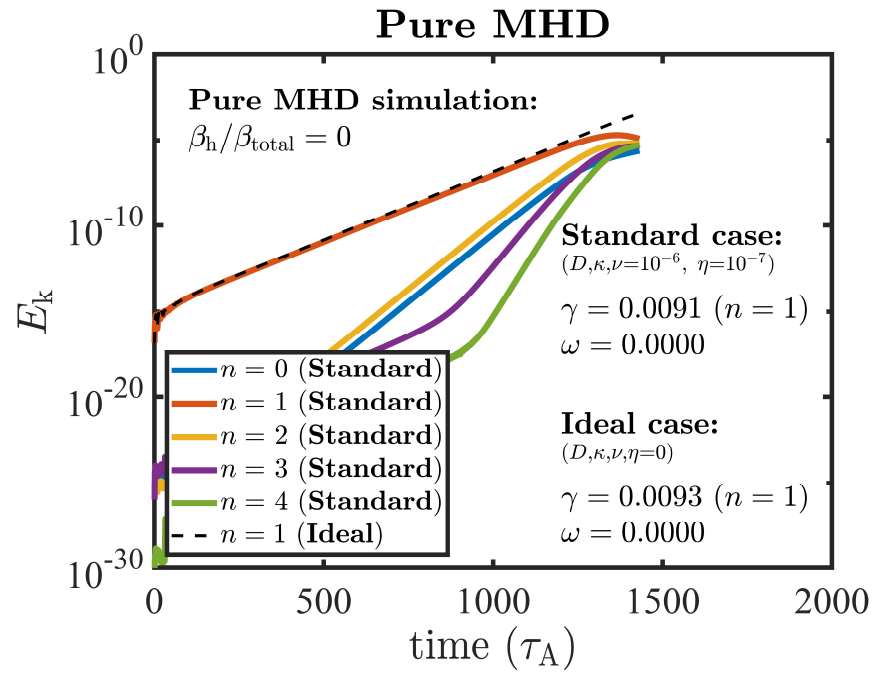

***Figure 2 :*** Evolution of the kinetic energy for IKMs in pure MHD simulations. Colored solid lines represent the standard case (with a small dissipation coefficient) for $n=0\sim4$, while the black dashed line represents the ideal case (no dissipation, filtering to retain only $n=1$).

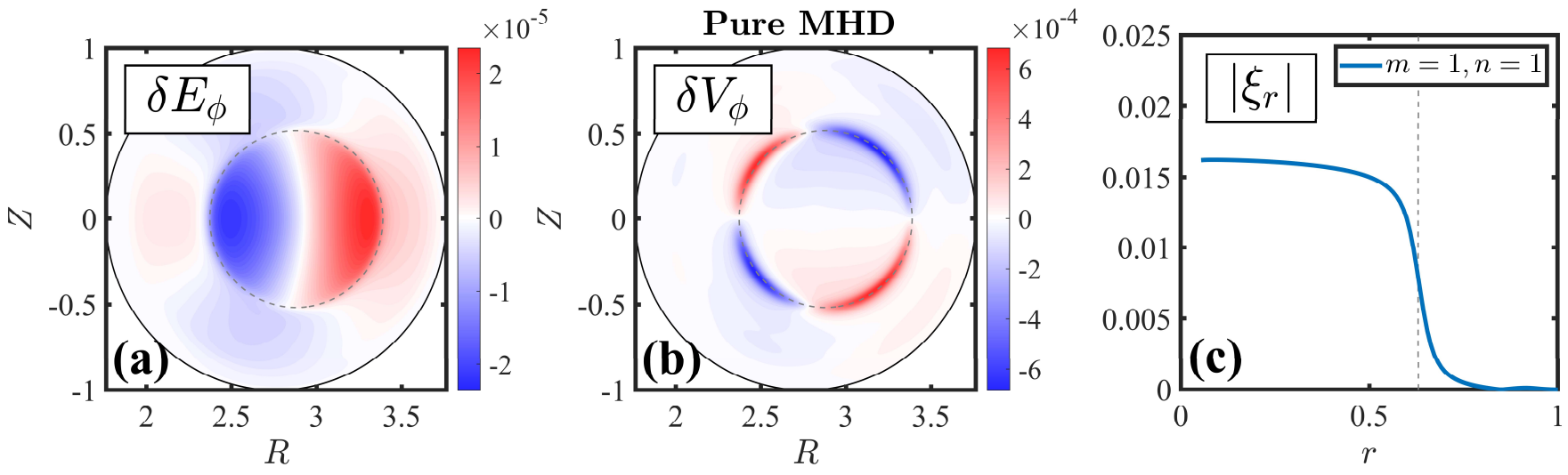

***Figure 3 :*** Mode structures at a typical moment during the linear phase of the IKM obtained in pure MHD simulations, represented by (a) the toroidal electric field $\delta E_\phi$, (b) the toroidal fluid velocity $\delta V_\phi$, and (c) the radial displacement $|\xi_r|$. (Due to the singularity of $|\xi_r|$ and numerical errors in the transformation between cylindrical and flux surface coordinates, calculations of $|\xi_r|$ become inaccurate very close to the magnetic axis, so the line is not extended into this region.) The gray dashed line represents the $q=1$ rational surface (also consistent in other figures of this paper).

In Figure 2, comparison results for two cases under ideal MHD conditions $(D,\kappa,\nu,\eta=0$, filtered to retain only the $n=1$ mode) and with weak dissipation coefficients $(D,\kappa,\nu=10^{-6}$, $\eta=10^{-7})$ indicate that the chosen weak dissipation coefficients have negligible effects on the linear results ($\gamma=0.0093$ versus $\gamma=0.0091$). For numerical stability consideration, we will use this set of weak dissipation coefficients in the other simulations presented in this paper.

## 4.3 Simulation Results with Nearly Isotropic EPs

Now we consider the existence of nearly isotropic EPs and use $\beta_h/\beta_{total}$ to measure the fraction of EPs within the total plasma. We take three cases of $\beta_h/\beta_{total}=0.25, 0.50, 0.75$.

At the end of Section 3, we discussed three different coupling schemes adopted in the simulations. The linear growth rates and mode frequencies of the IKMs are plotted in Figure 4. For comparison, the results obtained in Reference [9] using M3D-K and NOVA2, as well as the results obtained in Reference [13] and its subsequent works using M3D-C1-K, are also plotted in Figure 4 after being rescaled according to the definition of $\omega_A = v_A/a$. The mode frequency is defined as positive when rotating along the ion diamagnetic drift direction.

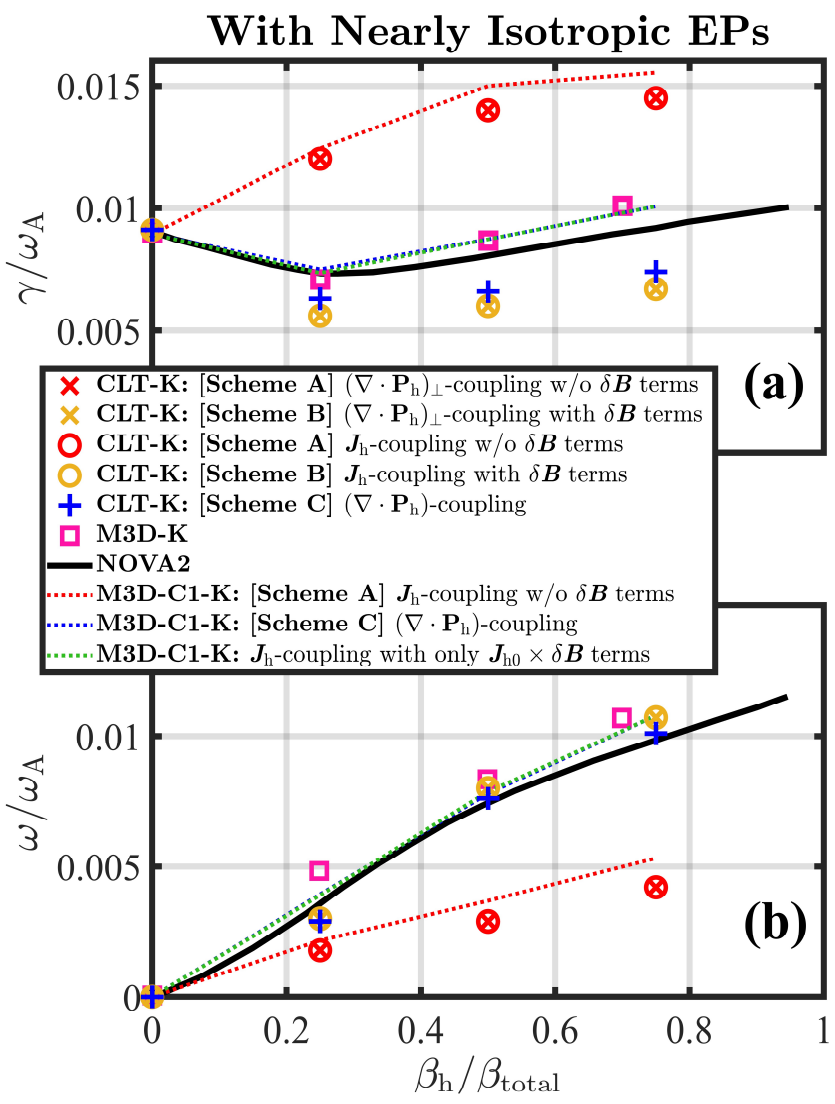

***Figure 4 :*** (a) Linear growth rates and (b) mode frequencies of the IKM in the presence of nearly isotropic EPs calculated using different coupling schemes, with results from CLT-K, M3D-K, NOVA2, and M3D-C1-K. (Please note the completely overlap between the red circles and red crosses, and between the yellow circles and yellow crosses, as well as the approximate overlap of yellow markers (circles and crosses) with blue plus signs, and the blue dotted line with the green dotted line.)

First, we discuss the case using Scheme C to benchmark against the results in Reference [9]. It can be observed that in this case, the effect of EPs on the IKM is primarily stabilizing (more precisely, as the fraction of EPs increases, the growth rate of the IKM first decreases and then increases, exhibiting a distinct turning point), and a higher EP fraction leads to a higher mode frequency. This result is quantitatively consistent with that from M3D-K and NOVA2. This shows that the two codes (CLT-K and M3D-K) can obtain basically consistent results under the same coupling scheme. Furthermore, the trend observed in the results from CLT-K and M3D-C1-K also demonstrates this consistency (for both Scheme A and Scheme C). The remaining minor differences, especially the slightly lower growth rate from CLT-K, are likely due to different numerical methods used by the different codes.

Secondly, the crosses (results of $(\nabla\cdot\mathbf{P}_h)_\perp$-coupling) and circles (results of current-coupling) of the same color completely overlap, demonstrating the reaffirmed equivalence of the pressure and current coupling schemes. However, it is evident that Scheme A leads to a significantly higher growth rate and a noticeably lower frequency of the IKM compared to the results from Scheme B.

This suggests that the introduction of the $\delta \boldsymbol{B}$ contribution has a significant stabilizing effect on the IKM. Scheme C also differs notably from the results from Scheme A, indicating that $(\nabla \cdot \delta \mathbf{P}_{\mathrm{h}})_{\parallel}^{(\delta f)}$ shows a stabilization effect. Interestingly, the results from Scheme B and Scheme C are very close, which suggests implicit connections between the stabilizing effects of these two terms and will be further discussed in Sections 4.4 and 4.5.

Taking the example of $\beta_{\mathrm{h}}/\beta_{\mathrm{total}} = 0.25$ (A smaller fraction of EPs allows the IKM to maintain MHD-like characteristics as much as possible), the time evolutions of the kinetic energy for the three coupling schemes are shown in Figure 5. This allows us to more intuitively observe both the previously mentioned equivalence and the significant effects caused by different schemes. Figure 6 also shows the typical linear phase mode structure for the IKM obtained using the three schemes. It should be noted that when in Scheme A, a strong $m/n = 1/1$ toroidal flow is excited on the inner side of the $q = 1$ rational surface, which may be responsible for making the mode more unstable, leading to a lower frequency. The same toroidal flow can also be observed in the simulations using Scheme A of M3D-C1-K, as presented in Appendix B. For Scheme B, Figure 7 also shows $\delta P_{\mathrm{h}\parallel}$, $\delta P_{\mathrm{h}\perp}$ and their difference, which is consistent with the results in Reference [13].

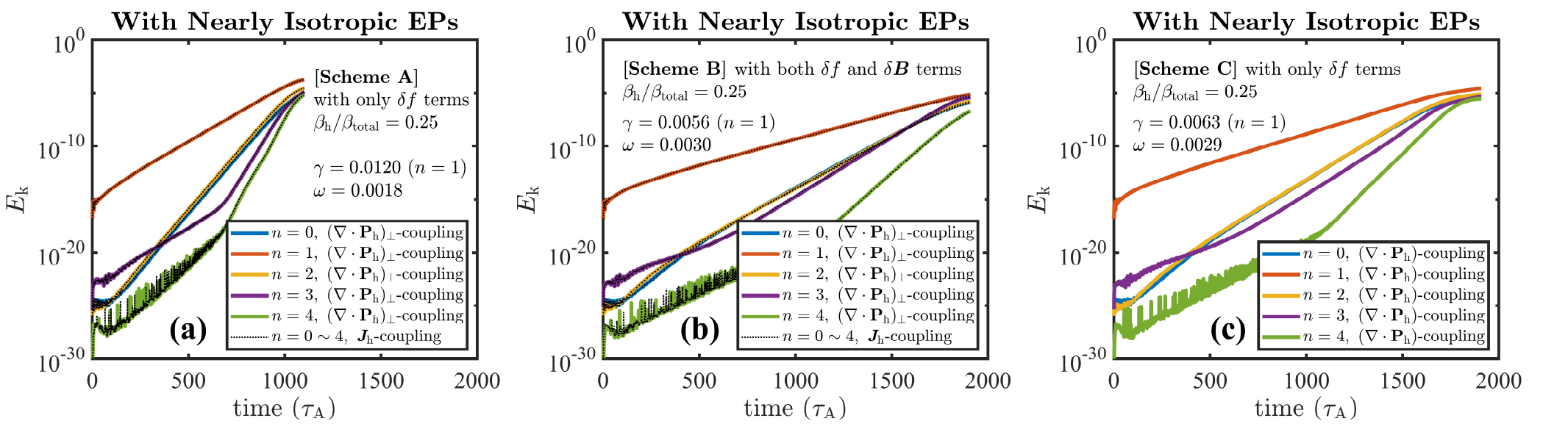

***Figure 5 :*** Evolution of the kinetic energy of the IKM obtained from (a) Scheme A, (b) Scheme B, and (c) Scheme C, with nearly isotropic EPs ($\beta_{\mathrm{h}}/\beta_{\mathrm{total}} = 0.25$). Colored solid lines represent the results using $(\nabla \cdot \mathbf{P}_{\mathrm{h}})_{\perp}$-coupling ($n = 0 \sim 4$), while the black dotted lines in (a) and (b) represent the results using current-coupling ($n = 0 \sim 4$), which are strictly equivalent.

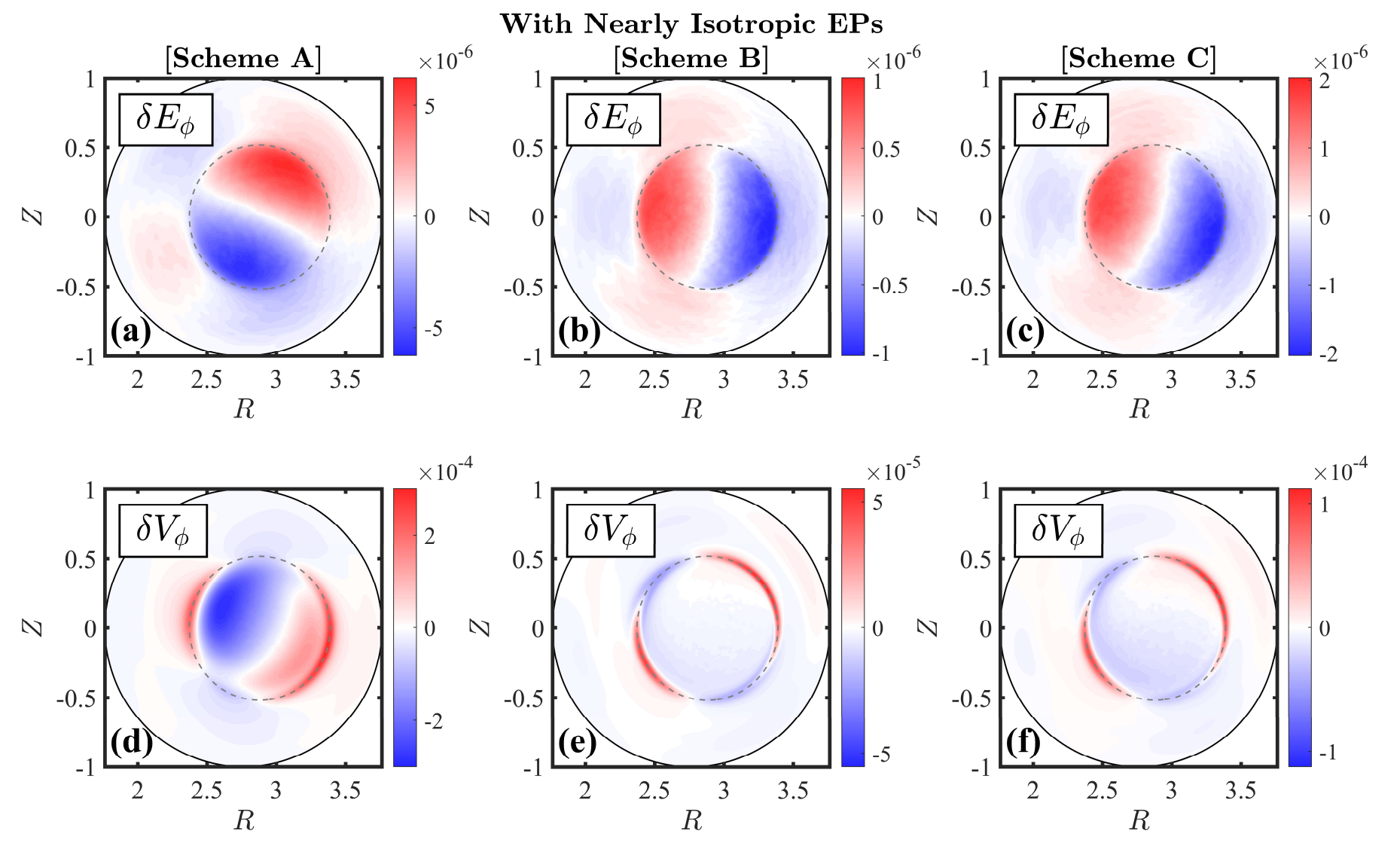

*Figure 6 :* Structures of (a~c) the toroidal electric field $\delta E_\phi$ and (d~f) the toroidal fluid velocity $\delta V_\phi$ at a typical moment during the linear phase of the IKM obtained from different coupling schemes with nearly isotropic EPs ($\beta_\mathrm{h}/\beta_\mathrm{total}=0.25$).

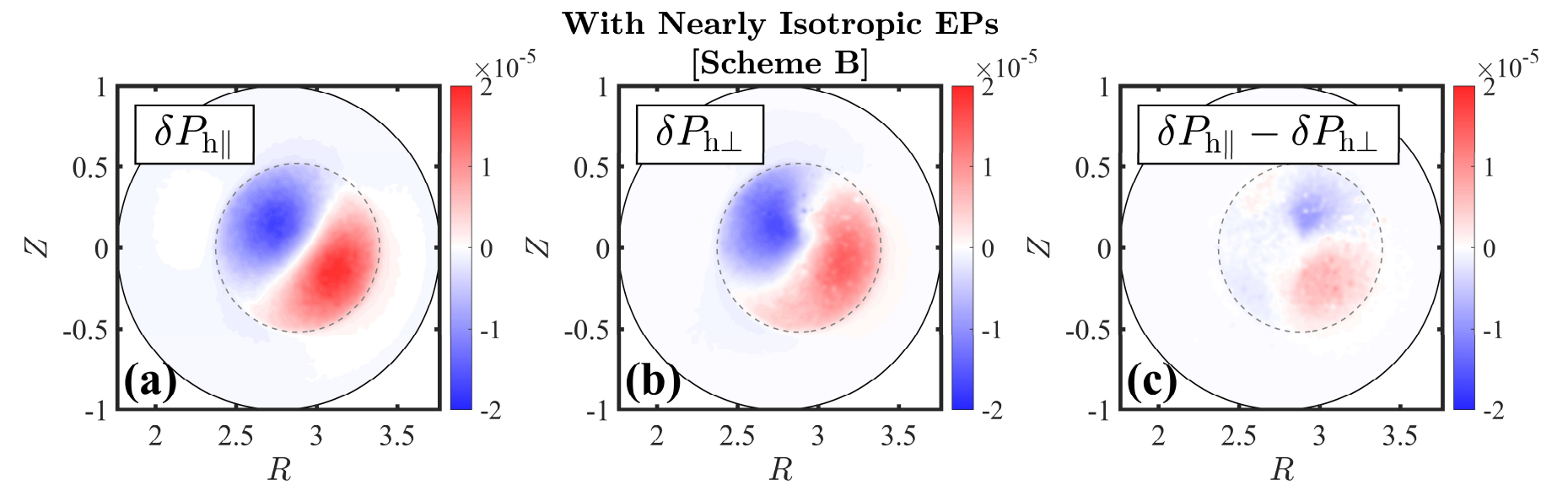

*Figure 7 :*(a) The parallel component, (b) the perpendicular component, and (c) their difference of the EP pressure at a typical moment during the linear phase of the IKM obtained from Scheme B with nearly isotropic EPs ($\beta_\mathrm{h}/\beta_\mathrm{total}=0.25$).

## 4.4 Energy Principle Analysis

From the above simulation results, we can conclude that the $\delta\boldsymbol{B}$ contribution and the $(\nabla\cdot\delta\mathrm{P_h})_\parallel^{(\delta f)}$ term both have significant stabilizing effects on the IKM. This suggests that in the hybrid simulations, the parallel inertia term of EPs and the $\delta\boldsymbol{B}$ contribution due to magnetic field perturbations should not be ignored, at least for the IKM cases. Thus, the more rigorous approach Scheme B should be adopted, that is, the $(\boldsymbol{\nabla}\cdot\mathbf{P}_\mathrm{h})_\perp$-coupling scheme (or equivalent current-coupling scheme) that contains both the $\delta f$ and $\delta\boldsymbol{B}$ contribution terms. Then it can return to Park's original model [1].

The energy principle is helpful in understanding the roles of these different terms. The linearized energy conservation equation can be written as:

$$\delta E_\mathrm{k} + \delta W_\mathrm{MHD} + \delta W_\mathrm{h} = 0 \tag{35}$$

where $\delta E_\mathrm{k} = \frac{1}{2}\int \rho|\dot{\boldsymbol{\xi}}|^2 \mathrm{d}V$ and $\delta W_\mathrm{MHD}$ are respectively the perturbation kinetic energy and potential energy of background plasma, and $\delta W_\mathrm{h}$ is the contribution of EPs to the perturbation potential energy. In the three different coupling schemes, $\delta W_\mathrm{h}$ can be expressed as:

$$\delta W_\mathrm{h}^{[A]} = \frac{1}{2}\int \boldsymbol{\xi}\cdot\left\{(\nabla\cdot\delta\mathbf{P}_\mathrm{h})_\perp^{(\delta f)}\right\}\mathrm{d}V \tag{36}$$

$$\delta W_\mathrm{h}^{[B]} = \frac{1}{2}\int \boldsymbol{\xi}\cdot\left\{(\nabla\cdot\delta\mathbf{P}_\mathrm{h})_\perp^{(\delta f)} + \delta[(\nabla\cdot\mathbf{P}_\mathrm{h})_\perp]^{(\delta\boldsymbol{B},\mathrm{perp})} + \delta[(\nabla\cdot\mathbf{P}_\mathrm{h})_\perp]^{(\delta\boldsymbol{B},\mathrm{cur})}\right\}\mathrm{d}V \tag{37}$$

$$\delta W_\mathrm{h}^{[C]} = \frac{1}{2}\int \boldsymbol{\xi}\cdot\left\{(\nabla\cdot\delta\mathbf{P}_\mathrm{h})_\perp^{(\delta f)} + (\nabla\cdot\delta\mathbf{P}_\mathrm{h})_\parallel^{(\delta f)}\right\}\mathrm{d}V \tag{38}$$

The superscript of $\delta W_\mathrm{h}$ is used to distinguish among different coupling schemes. By analyzing the sign of the integrated result of each term, the stabilizing or destabilizing effect of them can be judged.

Furthermore, in the $\delta f$ contribution of EPs, the adiabatic and non-adiabatic contributions can also be distinguished. The non-adiabatic contribution originates from the interaction between waves and EPs, and the adiabatic one is similar to the fluid-like contribution of the background plasma. Disregarding the effect of the finite orbit width, the adiabatic contribution can be approximated by $\delta f = \delta f_\mathrm{adi} \equiv -\boldsymbol{\xi}\cdot\nabla f_0$, and the term $(\nabla\cdot\delta\mathbf{P}_\mathrm{h})_\perp^{(\delta f,\mathrm{adi})}$ calculated in this way reflects the adiabatic contribution alone.

We select a typical moment in the linear phase from Scheme B and $\beta_\mathrm{h}/\beta_\mathrm{total} = 0.25$, integrate the kinetic and potential energy contributions of various terms in the EP contribution over the toroidal direction, and present the 2D distributions in Figure 8(b~g). The definitions of the specific potential terms can be seen in Table 1. We also calculate the total energy contributions by integrating them in the whole space (normalized by the total kinetic energy contribution at that moment). This provides a clear view of the stabilizing or destabilizing effect of each term. The evolution of each term contribution to the total energy over time is plotted in Figure 8(a). It can be found that the $\delta f$ contribution has a destabilizing effect, while the $\delta\boldsymbol{B}$ contribution has a stabilizing effect (for nearly isotropic distribution of particles, primarily from the "perp-terms"). The $(\nabla\cdot\delta\mathrm{P}_\mathrm{h})_\parallel^{(\delta f)}$ term subtracted from the $\delta f$ contribution also has a stabilizing effect that is very close to that of the $\delta\boldsymbol{B}$ contribution, as shown in Figure 8 (d) and (g). It explains why results from Scheme B and C are very similar. The above conclusion is consistent with the results obtained directly through simulations. Additionally, the adiabatic component of the $\delta f$ contribution significantly destabilizes, while the non-adiabatic component weakly stabilizes.

***Table 1 :*** **Definitions of each term in the energy principle analysis.**

| Terms | Definitions | Terms | Definitions |
|---|---|---|---|
| $\boldsymbol{\delta E_{\mathrm{k}}}$ | $\frac{1}{2}\int \rho\lvert\dot{\xi}\rvert^2 \mathrm{d}V$ | $\delta W_{\mathrm{h}}^{(\delta f,\perp)}$ | $\frac{1}{2}\int \xi\cdot\left\{(\nabla\cdot\delta\mathbf{P}_{\mathrm{h}})_{\perp}^{(\delta f)}\right\}\mathrm{d}V$ |
| $\boldsymbol{\delta W_{\mathrm{h}}^{(\delta B,\mathrm{perp})}}$ | $\frac{1}{2}\int \xi\cdot\{\delta[(\nabla\cdot\mathbf{P}_{\mathrm{h}})_{\perp}]^{(\delta\boldsymbol{B},\mathrm{perp})}\}\mathrm{d}V$ | $\delta W_{\mathrm{h}}^{(\delta B,\mathrm{cur})}$ | $\frac{1}{2}\int \xi\cdot\{\delta[(\nabla\cdot\mathbf{P}_{\mathrm{h}})_{\perp}]^{(\delta\boldsymbol{B},\mathrm{cur})}\}\mathrm{d}V$ |
| $\boldsymbol{\delta W_{\mathrm{h}}^{(\delta f,\mathrm{adi}\perp)}}$ | $\frac{1}{2}\int \xi\cdot\left\{(\nabla\cdot\delta\mathbf{P}_{\mathrm{h}})_{\perp}^{(\delta f,\mathrm{adi})}\right\}\mathrm{d}V$ | $\delta W_{\mathrm{h}}^{(\delta f,\parallel)}$ | $\frac{1}{2}\int \xi\cdot\left\{(\nabla\cdot\delta\mathbf{P}_{\mathrm{h}})_{\parallel}^{(\delta f)}\right\}\mathrm{d}V$ |

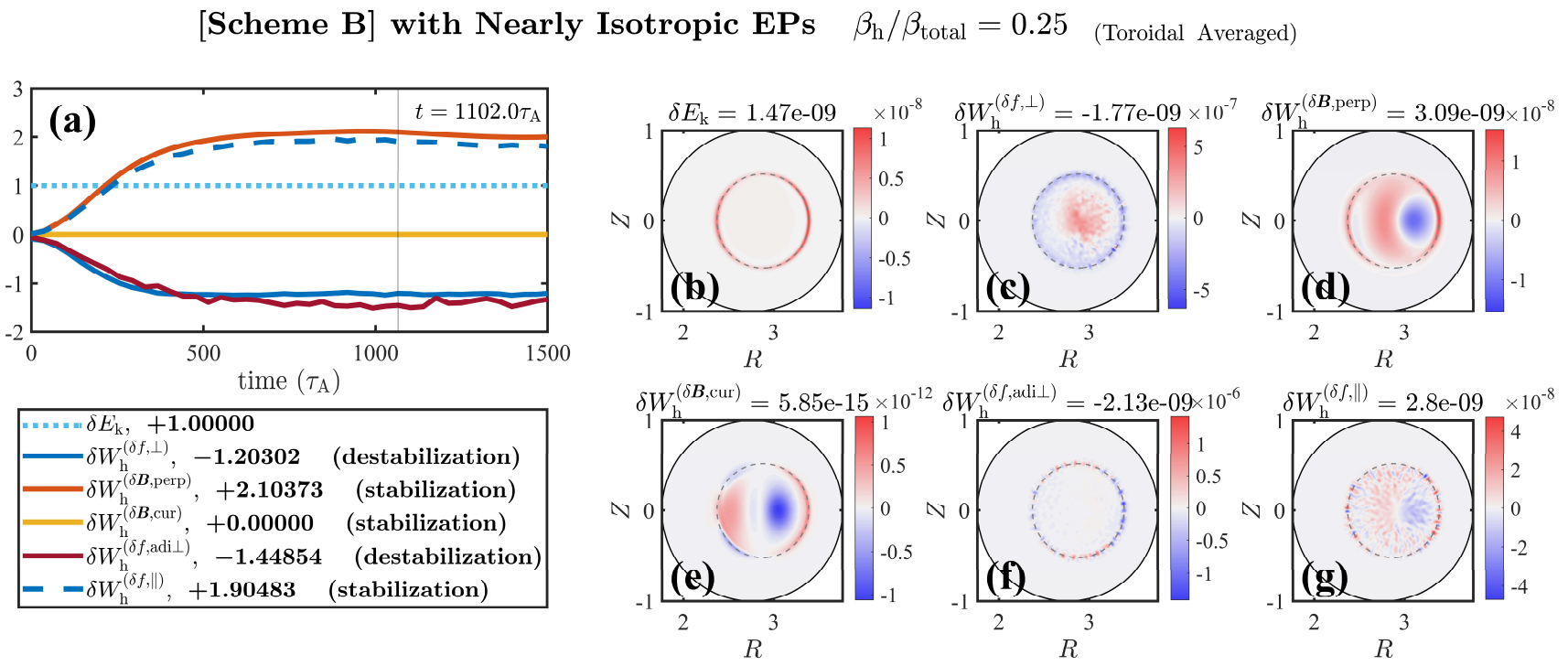

***Figure 8 :*** (a) Time evolutions of the total contributions of the kinetic and perturbed potential energy terms in Equations (36-38) using Scheme B with nearly isotropic EPs ($\beta_h/\beta_{total} = 0.25$) and (b~g) the 2D distribution of each term contributions at a typical moment (marked by the black vertical line in (a)). The 2D distribution is toroidally averaged. The total contributions in (a) and its legend are all normalized by $\delta E_k$ at the current moment.

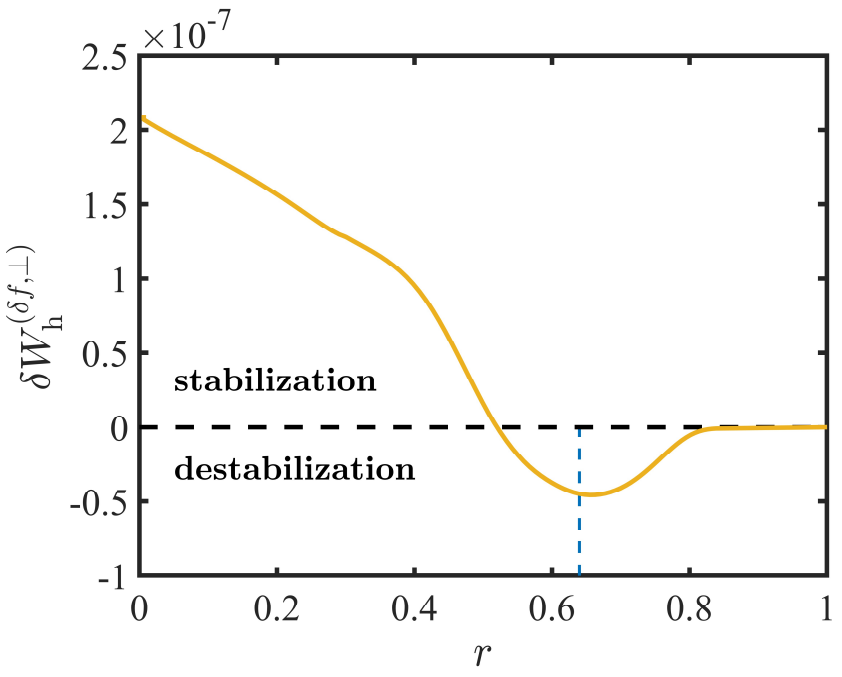

***Figure 9 :*** Radial distribution of the $\delta f$ contribution from Scheme B with nearly isotropic EPs ($\beta_h/\beta_{total} = 0.25$), where the curves have been smoothed for illustrative purposes. The blue dashed line indicates the position of the $q = 1$ rational surface.

Interestingly, according to Figure 8(c), the $\delta f$ contribution of EPs plays a destabilizing role primarily near the $q = 1$ rational surface and a stabilizing role primarily on the region inside of the $q = 1$ rational surface, which suggests that, for better stability, it is advisable to confine energetic particles inside the $q = 1$ rational surfaces, as depicted in Figure 9, which is in well agreement with the theoretical predictions by Porcelli et al. in References [35] and [36].

### 4.5 Inference Based on the Assumption of Isotropic EP Distribution

In the above study, we observe that for nearly isotropic distribution of EPs, the results obtained from Scheme B and C are very close, meaning the stabilizing effects of $\delta[(\nabla\cdot\mathbf{P}_\mathrm{h})_\perp]^{(\delta\boldsymbol{B},\mathrm{perp})}$ and $(\nabla\cdot\delta\mathrm{P}_\mathrm{h})_\parallel^{(\delta f)}$ are almost identical. In fact, this interesting conclusion can be approved under the condition that the initial EP pressure $\mathbf{P}_{\mathrm{h}0}$ is isotropically distributed.

Assuming $\mathbf{P}_{\mathrm{h}0}$ is strictly isotropic, i.e., $P_{\mathrm{h}\perp0}=P_{\mathrm{h}\parallel0}$, the difference between these two terms is:

$$
\begin{aligned}
&(\nabla\cdot\delta\mathrm{P}_\mathrm{h})_\parallel^{(\delta f)}-\delta[(\nabla\cdot\mathbf{P}_\mathrm{h})_\perp]^{(\delta\boldsymbol{B},\mathrm{perp})}\\
&=(\nabla\delta P_{\mathrm{h}\perp})_\parallel+[(\delta P_{\mathrm{h}\parallel}-\delta P_{\mathrm{h}\perp})(\nabla\cdot\boldsymbol{b})+\boldsymbol{b}\cdot\nabla(\delta P_{\mathrm{h}\parallel}-\delta P_{\mathrm{h}\perp})]\boldsymbol{b}\\
&\qquad\qquad-[(\nabla P_{\mathrm{h}\perp0})_\perp-(\nabla P_{\mathrm{h}\perp0})_{\perp0}]\\
&=(\nabla\delta P_{\mathrm{h}\perp})_\parallel+(\nabla P_{\mathrm{h}\perp0})_\parallel+[(P_{\mathrm{h}\parallel}-P_{\mathrm{h}\perp})(\nabla\cdot\boldsymbol{b})+\boldsymbol{b}\cdot\nabla(P_{\mathrm{h}\parallel}-P_{\mathrm{h}\perp})]\boldsymbol{b}\\
&=(\nabla\cdot\mathrm{P}_\mathrm{h})_\parallel
\end{aligned}
\tag{39}
$$

In the case where the parallel inertia term of EPs is considered negligible ($\partial(\rho_\mathrm{h}\boldsymbol{V}_{\mathrm{h}\parallel})/\partial t=\mathbf{0}$) to the extent that $(\nabla\cdot\mathbf{P}_\mathrm{h})$ is perpendicular to $\boldsymbol{b}$, i.e., $(\nabla\cdot\mathbf{P}_\mathrm{h})_\parallel=\mathbf{0}$, the terms $\delta[(\nabla\cdot\mathbf{P}_\mathrm{h})_\perp]^{(\delta\boldsymbol{B},\mathrm{perp})}$ and $(\nabla\cdot\delta\mathrm{P}_\mathrm{h})_\parallel^{(\delta f)}$ are approximately equal [d]. This explains why the results from Scheme B and C are nearly the same and why many pressure-coupling codes [11,13] using Scheme B and C could produce similar results. In the simulations using M3D-C1-K from Reference [13], the current-coupling scheme that includes a partial $\delta\boldsymbol{B}$ contribution [e] and $(\nabla\cdot\mathbf{P}_\mathrm{h})$-coupling scheme provide similar results (also illustrated in Figure 4 of this paper, showing the approximate overlap of the blue and green dotted lines), again for this reason.

However, it should be emphasized that this approximated inference holds only when EPs exhibit isotropic distribution. The formulation of Scheme B is more generally applicable. The next section will analyze the case of anisotropic distribution of EPs.

[d] Note that Scheme C only neglects a part of the EP parallel inertial term, not the entire term. $(\nabla\cdot\delta\mathrm{P}_\mathrm{h})_\parallel^{(\delta f)}$ includes only the $\delta f$ contribution from $(\nabla\cdot\mathbf{P}_\mathrm{h})_\parallel$.

[e] In this reference, the $\delta\boldsymbol{B}$ contribution in the current-coupling scheme only includes $\boldsymbol{J}_{\mathrm{h}0}\times\delta\boldsymbol{B}=-\left(\nabla p_{\mathrm{h}0}\cdot\frac{\delta\boldsymbol{B}}{B_0}\right)\boldsymbol{b}_0$, which is close to the complete form $-(\nabla p_{\mathrm{h}0}\cdot\delta\boldsymbol{b})\boldsymbol{b}$. Here, $p_{\mathrm{h}0}$ is the initial scalar pressure of isotropic EPs. These results are also plotted in Figure 4 using green dotted lines.

# 5. Influences of Anisotropic EPs on the Stability of the 1/1 IKM

## 5.1 Initial Equilibrium and Simulation Parameters

In this section, we will continue discussing the cases anisotropic EP distributions. We still use the equilibrium and parameters consistent with Section 3 of Reference [9]. The only difference is that the EP distribution function with different pitch angle $\Lambda$ is considered:

$$f_0 = \frac{1}{v^3 + v_c^3}\left[1 + \mathrm{erf}\left(\frac{v_0 - v}{\Delta v}\right)\right] \exp\left(-\frac{\overline{\langle\psi\rangle}}{0.25}\right) \exp\left[-\left(\frac{\Lambda - \Lambda_0}{\Delta\Lambda}\right)^2\right] \tag{40}$$

where $\Delta\Lambda = 0.25$; for cases where trapped particles dominate, $\Lambda_0 = 1.0$; for cases where co-passing or counter-passing particles dominate, $\Lambda_0 = 0.0$. All other settings remain identical to Section 4.1.

It's important to note that this paper assumes that the background plasma and EPs jointly contribute to the equilibrium pressure, satisfying the Equation (12). However, since $\mathbf{P}_{\mathrm{h0}}$ cannot reduce to a scalar for the anisotropic EP distribution, the thermal equilibrium may not be accurate (this issue has been discussed in Section 2.3). [f]

## 5.2 Simulation Results with Anisotropic EP Distribution

We examined three cases: trapped particle-dominated, co-passing particle-dominated, and counter-passing particle-dominated. The calculated linear growth rates and frequencies for different schemes and different $\beta_{\mathrm{h}}$ are shown in Figure 10. Obviously, different coupling schemes still bring significant differences that also largely differ from that of nearly isotropic EP distribution. In the cases of anisotropic EP distribution, the results from Schemes B and C are no longer close, and the difference between the results from Schemes A and C decreases. The mode structures from Scheme B are shown in Figure 11.

[f] Furthermore, the definition of $\beta_{\mathrm{h}}$ is also controversial, because different $\beta_{\mathrm{h}\parallel}$ and $\beta_{\mathrm{h}\perp}$ will be defined according to $P_{\mathrm{h}\parallel 0}$ and $P_{\mathrm{h}\perp 0}$ respectively. One approach is to define $\beta_{\mathrm{h}}$ as the average of $\beta_{\mathrm{h}\parallel}$ and $\beta_{\mathrm{h}\perp}$ (i.e., $\beta_{\mathrm{h}} \equiv (\beta_{\mathrm{h}\parallel} + \beta_{\mathrm{h}\perp})/2$). This approach may overestimate the influence of EPs. Another approach is to define $\beta_{\mathrm{h}}$ as one of the primary components between $\beta_{\mathrm{h}\parallel}$ and $\beta_{\mathrm{h}\perp}$ (i.e., $\beta_{\mathrm{h}} \equiv \max(\beta_{\mathrm{h}\parallel}, \beta_{\mathrm{h}\perp})$). This approach may underestimate the influence of EPs. In this paper, except for the cases in Section 5.4 that follow the former definition, all other cases adhere to the latter definition.

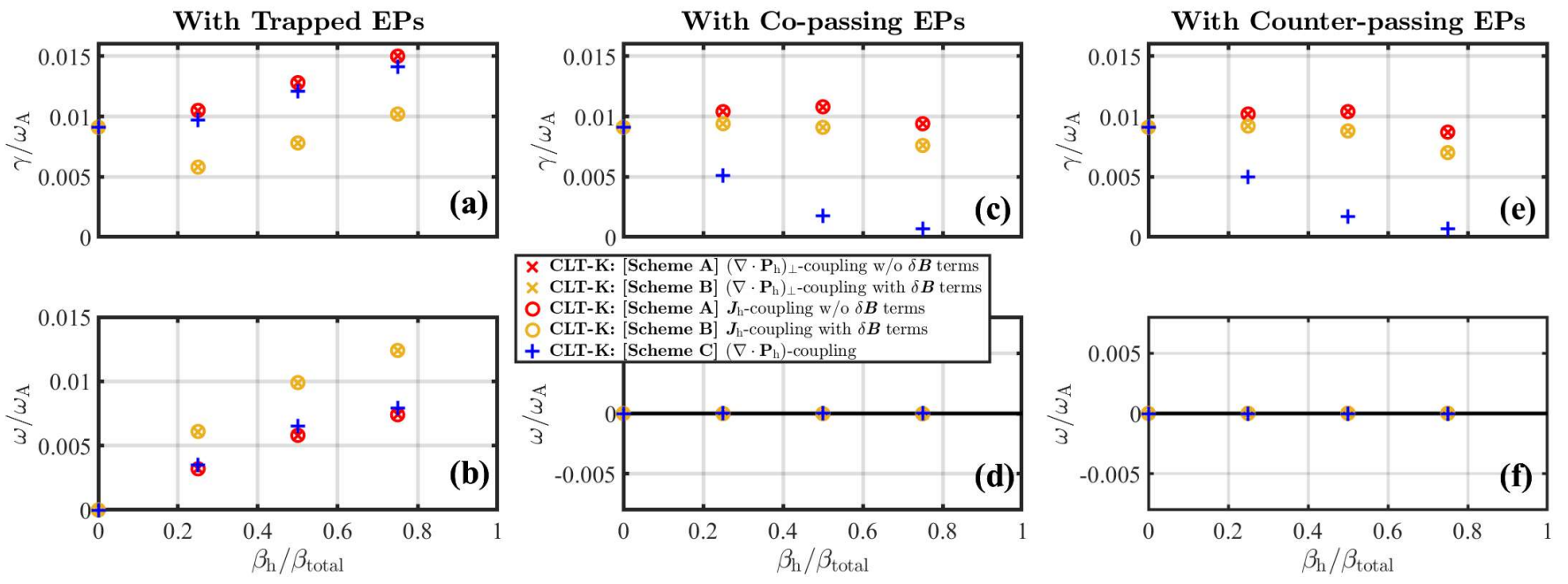

***Figure 10 :*** Linear growth rates (a,c,e) and mode frequencies (b,d,f) of the IKM in the presence of (a,b) trapped, (c,d) co-passing, and (e,f) counter-passing EPs calculated through different coupling schemes using CLT-K.

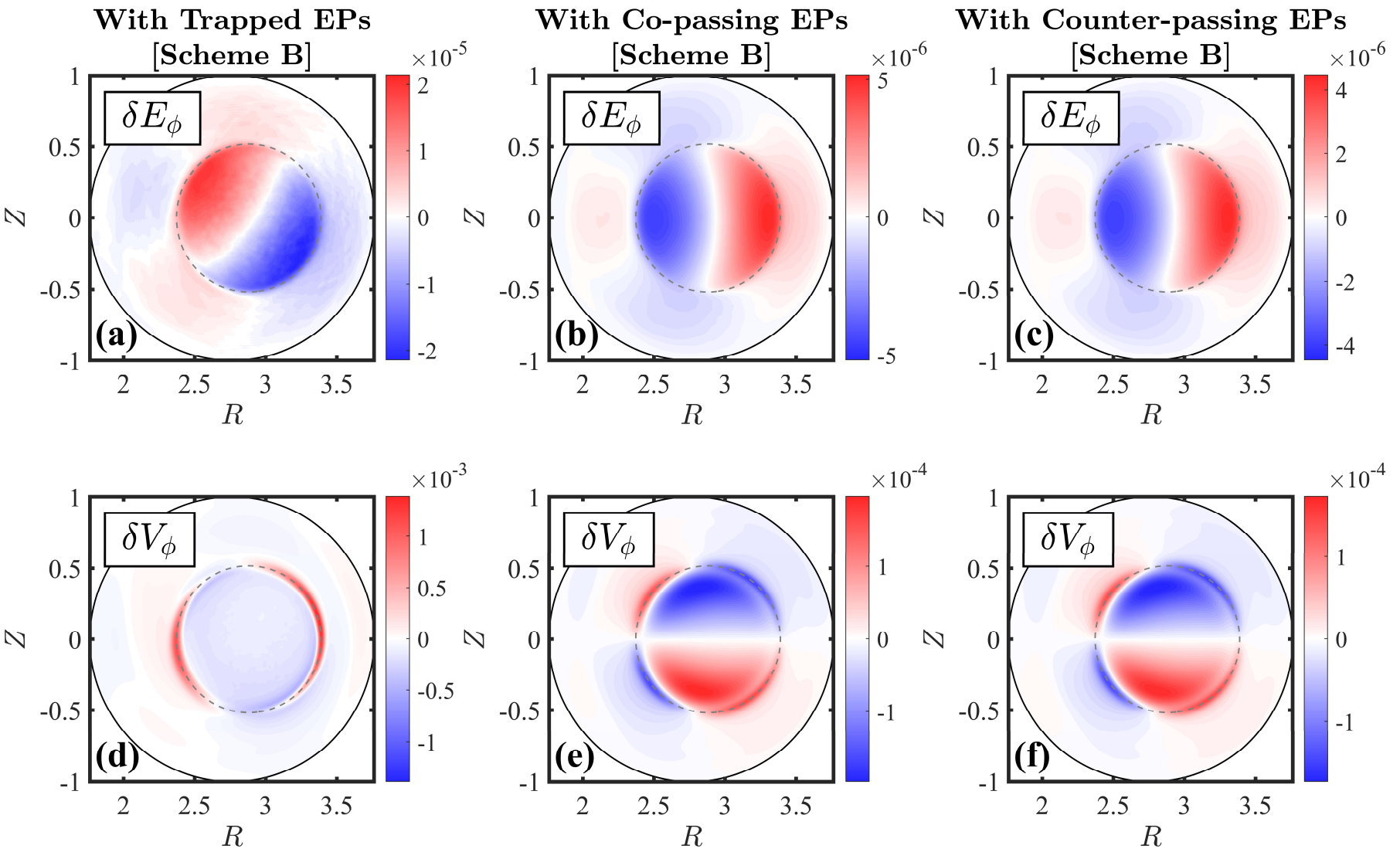

***Figure 11 :*** Structures of (a~c) the toroidal electric field $\delta E_\phi$ and (d~f) the toroidal fluid velocity $\delta V_\phi$ at a typical moment during the linear phase of the IKM from Scheme B with anisotropic EPs ($\beta_h/\beta_{total} = 0.25$).

Specifically, for the case dominated by trapped particles, compared with the simplest Scheme A, the growth rate from Scheme B is significantly lower and the frequency significantly increases, indicating that the $\delta\boldsymbol{B}$ contribution still plays a stabilizing role in general. The growth rate from Scheme C is slightly lower than that from Scheme A, and the frequency is slightly higher, which shows that the additional $(\nabla\cdot\delta\mathrm{P}_h)_\parallel^{(\delta f)}$ term provides a weak stabilizing effect. In the case with Scheme B, strong toroidal flow is not generated in the core region.

If passing particles dominate, the situation will be different. The first thing to notice is that when the passing particles dominate, the mode frequencies are very low, even approaching zero, while the growth rates are not significantly different from the pure MHD cases, so the influence of passing particles on the IKM frequency and growth rate is not significant. The $\delta\boldsymbol{B}$ contribution still shows a stabilizing effect, albeit weaker, which reduces the difference between Schemes A and

B. However, the difference between Schemes A and C is significant, because when the additional neglect of EP parallel inertia term is introduced, the $(\nabla \cdot \delta \mathbf{P}_{\mathrm{h}})_{\parallel}^{(\delta f)}$ term will show a stronger stabilizing effect, even making the IKM to approach stable at higher EP fractions. This once again illustrates that in cases where passing particles dominate, the difference between $(\boldsymbol{\nabla} \cdot \mathbf{P}_{\mathrm{h}})_{\perp}$-coupling and $(\boldsymbol{\nabla} \cdot \mathbf{P}_{\mathrm{h}})$-coupling cannot be ignored. In this case, using Scheme B leads to the generation of strong toroidal flow in the core region. Further demonstrations of this toroidal flow can be found in Appendix B.

In the most reasonable cases of Scheme B, when the fraction of EPs is not too large, trapped particles generally stabilize the IKM, while passing particles exhibit a weak destabilizing effect on IKM, which is consistent with theoretical expectations, such as the conclusions of White et al. [37] and Wu et al. [38]. In contrast, it should be highly noted that adopting the simplified Scheme A or Scheme C leads to completely opposite and unreasonable conclusions.

## 5.3 Energy Principle Analysis

To further understand the specific contributions of different terms, we analyze the results based on the energy principle using the same method as in Section 4.4. We also select the calculation example of Scheme B with $\beta_{\mathrm{h}}/\beta_{\mathrm{total}} = 0.25$. Both 2D distribution of energy contributions and whole-volume integrated value (normalized with kinetic energy) over time for each member in Table 1 are plotted. For the trapped particle case, the results are shown in Figure 12. For the co-passing particle case, the results are shown in Figure 13 (counter-passing particle cases are nearly identical).

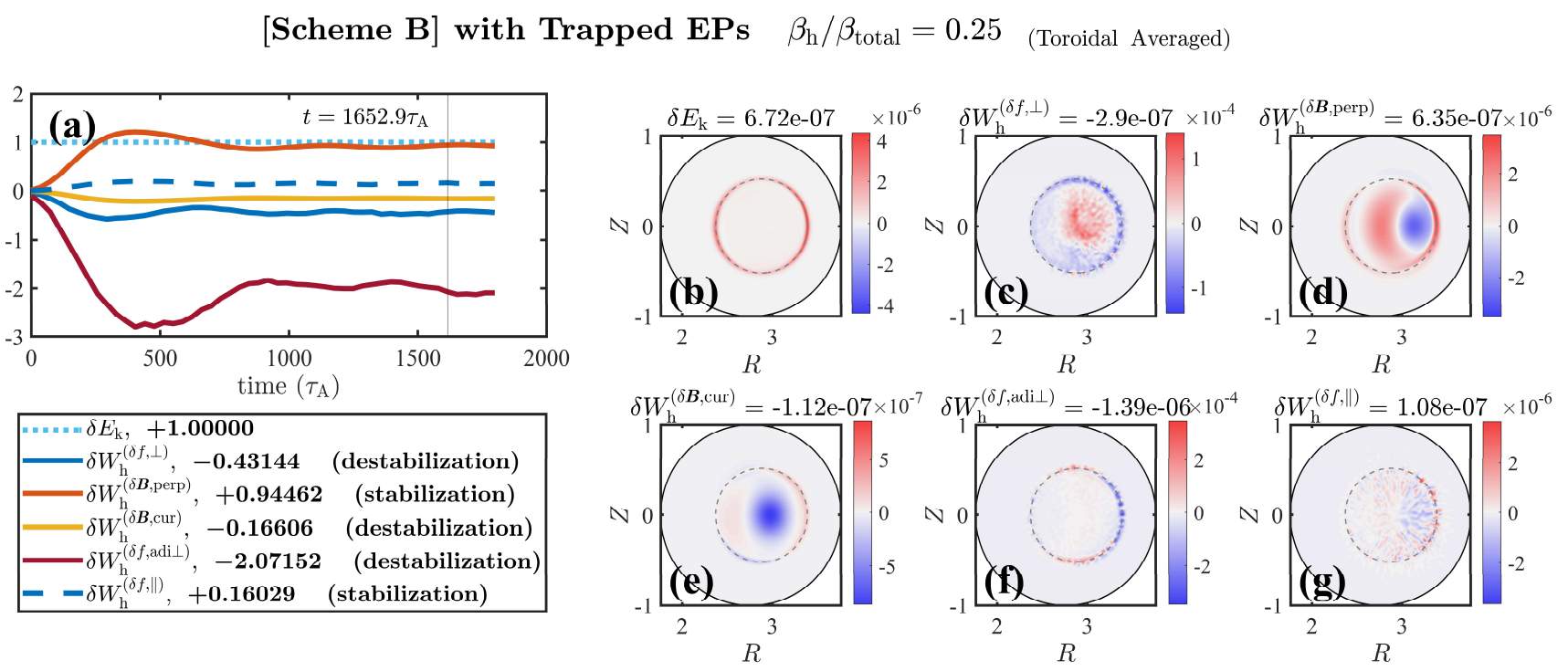

***Figure 12 :*** (a) Time evolutions of the total contributions of the kinetic and perturbed potential energy terms in Equations (36-38) using Scheme B with trapped EPs ($\beta_{\mathrm{h}}/\beta_{\mathrm{total}} = 0.25$) and (b~g) the 2D distribution of each term contribution at a typical moment (marked by the black vertical line in (a)). The 2D distribution is toroidally averaged. The total contributions in (a) and its legend are all normalized by $\delta E_{\mathrm{k}}$ at the current moment.

When trapped particles dominate, the $\delta \boldsymbol{B}$ contribution has an overall stabilizing effect, while the "perp-terms" and "cur-terms" respectively have a stabilizing and destabilizing effect, with the strength ratio of approximately 6:1. The $\delta f$ contribution still has a destabilizing effect. In detail, the parallel terms $(\nabla \cdot \delta \mathrm{P_h})_{\parallel}^{(\delta f)}$, subtracted in the $\delta f$ contribution, has only a very weak stabilizing effect. This is because trapped particles do not have a strong parallel inertia. The adiabatic term of the $\delta f$ contribution has a destabilizing effect, while the non-adiabatic term has a stabilizing effect. This conclusion qualitatively aligns with the theoretical results of Wu et al. [38] and Porcelli et al. [39].

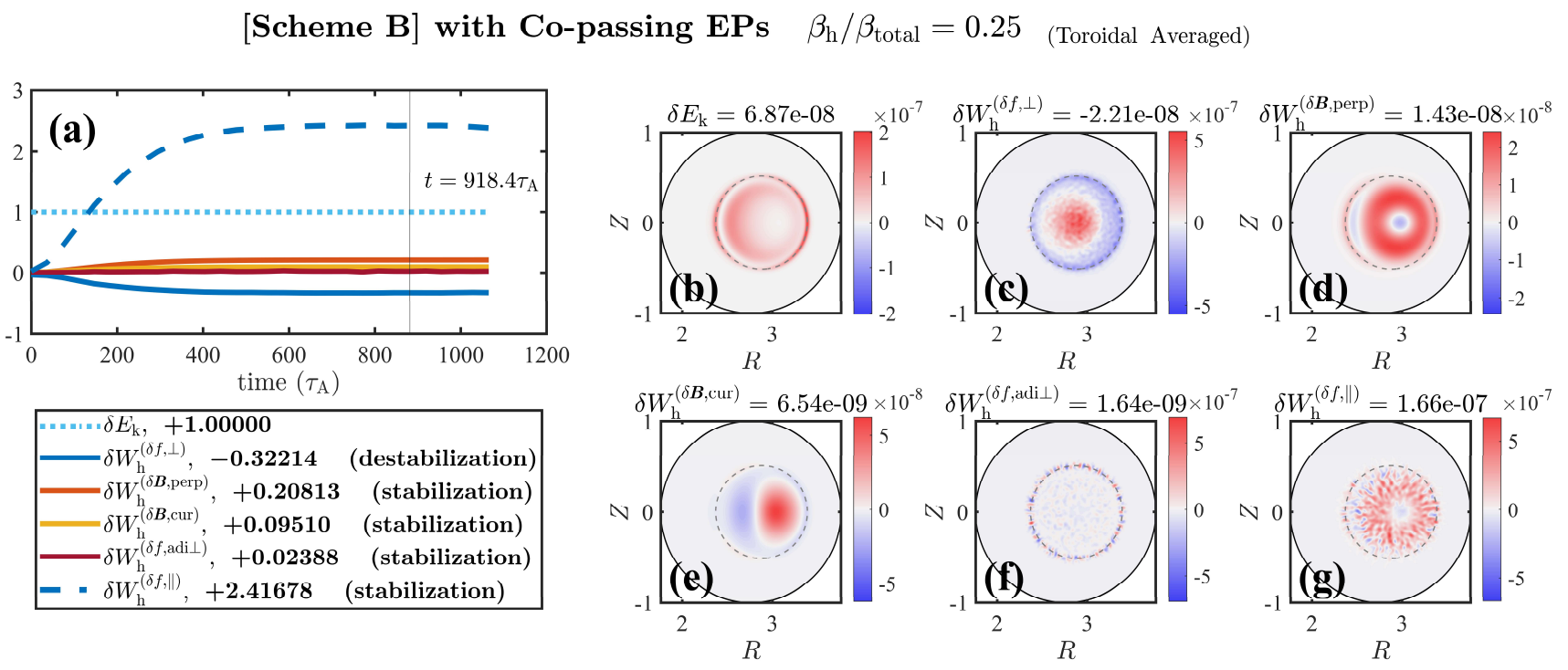

***Figure 13 :*** (a) Time evolutions of the total contributions of the kinetic and perturbed potential energy terms in Equations (36-38) from Scheme B with co-passing EPs ($\beta_h/\beta_{total} = 0.25$) and (b~g) the 2D distribution of each term contributions at a typical moment (marked by the black vertical line in (a)). The 2D distribution is toroidally averaged. The total contributions in (a) and its legend are all normalized by $\delta E_k$ at the current moment.

When co-passing particles dominate, the toroidal flow in the core region is excited. The $\delta \boldsymbol{B}$ contributions by both the "perp-terms" and "cur-terms" are stabilizing (with a ratio of 2:1), while the $\delta f$ contribution remains destabilizing. Past theoretical studies [38-41] also suggested that the adiabatic contribution of passing particles has a stabilizing effect, while the non-adiabatic contributions are weaker and depend on the injection direction of particles (if resonance occurs). Our results partially support this view: the adiabatic term is weakly stabilizing, while the non-adiabatic part will slightly destabilize the IKM. And the low mode frequency, independent of the particle injection direction, may serve as evidence that the influence of wave-particle resonance on the mode is not prominent in this case. The contribution of passing particles to the parallel component of pressure perturbation is very significant, making the $(\nabla \cdot \delta \mathrm{P_h})_{\parallel}^{(\delta f)}$ terms strongly stabilizing.

In conclusion, the analysis based on the energy principle aligns with the simulation results. With anisotropic particle distribution, the quantitative similarity between the respective contributions of the $\delta \boldsymbol{B}$ terms and the $(\nabla \cdot \delta \mathrm{P_h})_{\parallel}^{(\delta f)}$ term also breaks down.

## 5.4 Scanning the EP Distribution Function in Pitch Angle Space

Additionally, to clearly show the differences of various schemes under the EP anisotropic distribution, we scanned the EP distribution function in pitch angle space using the case of $\beta_h/\beta_{total} = 0.25$. We fixed $\Delta\Lambda$ at 0.25 and set $\Lambda_0$ to 0.0, 0.25, 0.5, 0.75, 0.9, 1.0, and 1.1, representing different scenarios from fully passing particles to deeply trapped particles. The simulation results are presented in Figure 14. [g]

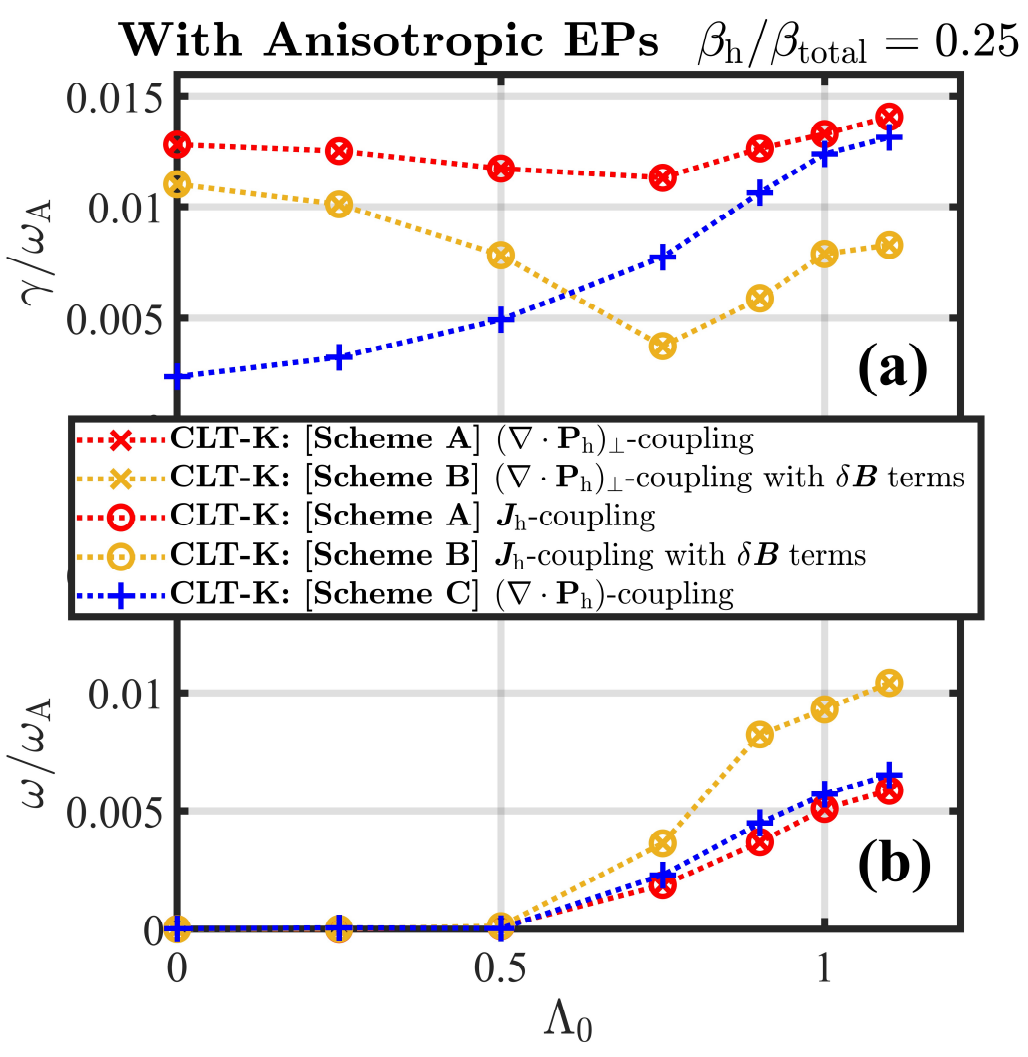

***Figure 14 :*** (a) Linear growth rates and (b) mode frequencies of the IKM with different coupling schemes using the CLT-K code, under varying anisotropic EPs in pitch angle space ($\beta_h/\beta_{total} = 0.25$ and $\Delta\Lambda$ is fixed at 0.25). Dotted lines indicate trends.

Figure 14 further clarifies our previous conclusions: when trapped particles dominate, Schemes A and C yield similar results, while Scheme B shows lower growth rates and higher mode frequencies. When passing particles dominate, the mode frequencies approach zero, and the growth rates of Schemes A and B become closer. As the proportion of passing particles increases, the stabilizing effect of the $(\nabla \cdot \delta \mathbf{P}_h)_\parallel^{(\delta f)}$ term gradually strengthens. The stronger the anisotropy of the initial distribution of the EPs (i.e., the greater the difference between $P_{h\perp 0}$ and $P_{h\parallel 0}$), the larger the difference in results between Schemes B and C.

[g] To maintain continuity for comparison across all cases, these cases differ from the others in this paper in two aspects: (1) The definition of $\beta_h$ follows $\beta_h \equiv (\beta_{h\parallel} + \beta_{h\perp})/2$ rather than the $\beta_h \equiv \max(\beta_{h\parallel}, \beta_{h\perp})$ used in other cases, as explained in footnote [f]. This may slightly overestimate the EP contribution under strong anisotropy but only affects the results quantitatively, not qualitatively. (2) For passing EPs, we no longer distinguish between co-passing and counter-passing EPs, as previous studies have shown that the direction of passing EPs has minimal effect on the results.

## 6. Influences of EP Larmor Radius

The typical Larmor radius of EPs is $\varrho_{\rm h} = 0.0125a$ (originally from Fu et al. [9]) in above simulations, which is relatively small. In this section, we will preliminarily investigate the influence of different $\varrho_{\rm h}$ on the results.

Still using Scheme B with $\beta_{\rm h}/\beta_{\rm total} = 0.25$ as an example, the simulation results with increasing $\varrho_{\rm h}$ for cases of particle isotropy, dominant trapped particles, dominant co-passing particles, and dominant counter-passing particles are shown in Figure 15.

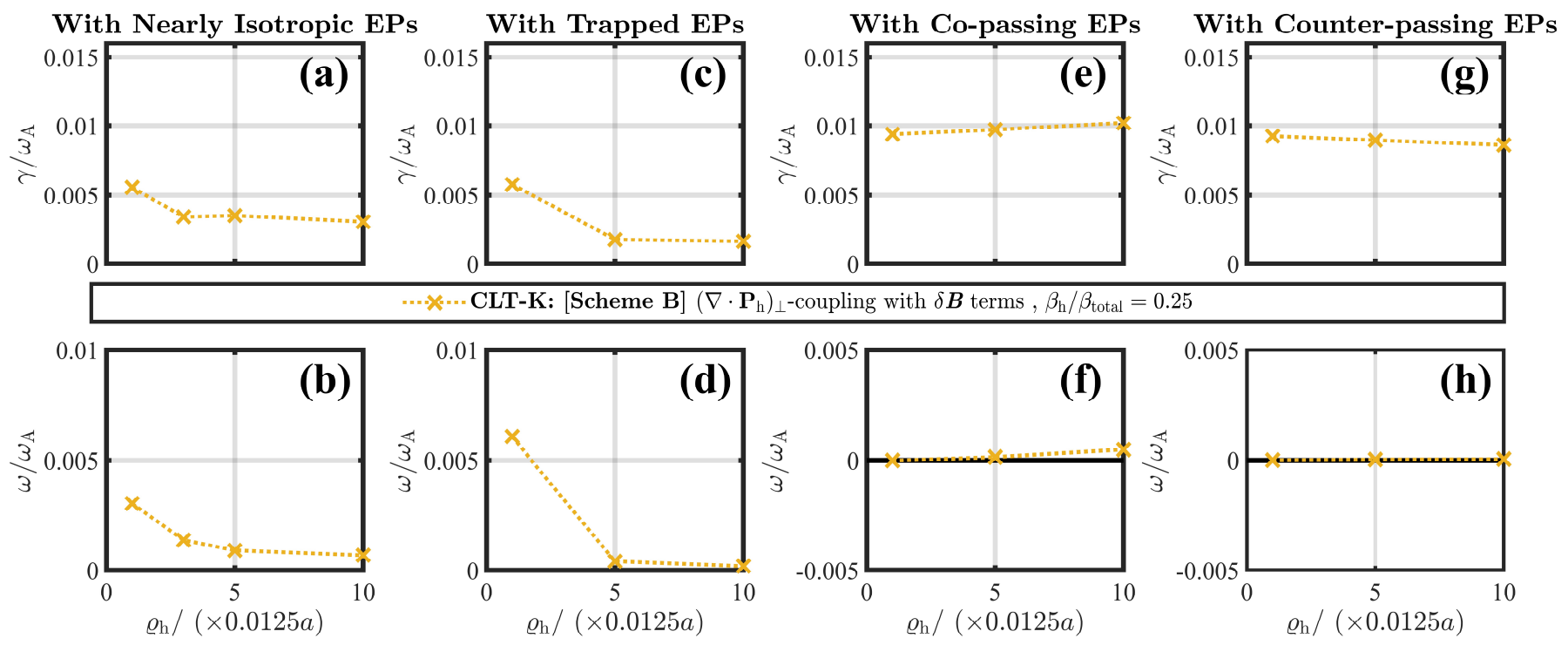

***Figure 15 :*** Linear growth rates (a,c,e,g) and mode frequencies (b,d,f,h) of the IKM in the presence of (a,b) nearly isotropic, (c,d) trapped, (e,f) co-passing, and (g,h) counter-passing EPs with different $\varrho_{\rm h}$ using CLT-K. Dotted lines indicate the trend.

It can be seen that with increasing $\varrho_{\rm h}$, there is no a significant change in the results of passing particles, which also reaffirms that the influence of passing particles on this IKM's frequency and growth rate is very limited. However, for the case dominated by trapped particles, the mode frequency rapidly decreases to near-zero levels with increasing $\varrho_{\rm h}$ (similar to passing particles), and the growth rate significantly decreases to a lower, nearly constant level (approximately $\gamma = 0.0017$). For the case with nearly isotropic EP distribution, the mode frequency decreases to near-zero levels, and the growth rate drops to approximately $\gamma = 0.0030$ that is a combination of the growth rate for passing and trapped EPs.

Since the simulations do not use a gyrokinetic model, the influences of $\varrho_{\rm h}$ are mainly reflected in changing the EP charge-to-mass ratio and orbit width. For trapped particles, increasing $\varrho_{\rm h}$ is akin to considering a more pronounced finite banana width (FBW) effect that leads to stabilize the IKM further, which is consistent with theoretical predictions by Helander et al. in Reference [42].

## 7. Summary and Discussion

In this study, we conducted a comprehensive analysis of different coupling schemes and methods in MHD-kinetic hybrid simulations, with a particular focus on the issues of the $\delta\boldsymbol{B}$ contribution and the neglect of parallel inertial terms. Using the CLT-K code, we investigated the influence of these different coupling schemes on the linear stability of IKMs in the presence of EPs. Our key findings can be summarized as follows: For IKMs cases with EPs, hybrid simulations should adopt the $(\nabla \cdot \mathbf{P}_{\mathrm{h}})_{\perp}$-coupling (or its equivalent current-coupling) scheme rather than the $(\nabla \cdot \mathbf{P}_{\mathrm{h}})$-coupling. Both the $\delta f$ and $\delta\boldsymbol{B}$ contributions of the same order should be included completely, that is, Scheme B should be adopted. This approach not only aligns with the original equations proposed by Park et al., but also ensures that pressure-coupling and current-coupling are strictly equivalent, regardless of the distribution of EPs. This marks the first validation of such equivalence in hybrid codes, provided that the forms of Equations (13) ~ (15) and (17) ~ (18) are adopted.

In our IKM simulations, the choice of whether to include the $\delta\boldsymbol{B}$ contribution and whether to use $(\nabla \cdot \mathbf{P}_{\mathrm{h}})_{\perp}$ -coupling or $(\nabla \cdot \mathbf{P}_{\mathrm{h}})$ -coupling significantly affects the simulation results. Specifically, the $\delta\boldsymbol{B}$ contribution introduces a noticeable stabilizing effect, while ignoring the parallel term $\delta[(\nabla \cdot \mathbf{P}_{\mathrm{h}})_{\parallel}]^{(\delta f)}$ may bring additional strong stabilization when passing particles dominate. Analysis based on the energy principle confirms the above conclusion. Assuming isotropy in the initial distribution of EPs and completely neglecting the parallel inertia terms of EPs, the stabilizing effects of these two terms are equivalent. This explains why codes using Scheme C can achieve results that are similar to Scheme B. However, this similarity may not hold universally, especially in cases with anisotropic EP distributions. The effects of specific terms on IKM obtained from the simulation are listed in Table 2. Note that these are only guaranteed to be valid for the parameters in this case. We also discussed the influences of increasing $\varrho_{\mathrm{h}}$ on the results, which suggests the strong FBW effect of trapped particles on influencing the stability of IKM.

We suggest that future research should pay further attention to the differences that different coupling schemes may introduce under different circumstances and the specific roles of each contribution term. This will help reevaluate existing results and improve the reliability of MHD-kinetic hybrid simulations. The introduction of $\delta\boldsymbol{B}$ contribution and the modification of the pressure-coupling model may be topics worthy of attention, particularly in studies investigating the influence of EPs on low-frequency MHD instabilities such as IKMs and TMs. Furthermore, it addresses the issues associated with a mechanical imbalance in cases with anisotropic EP distribution and the ambiguity in defining $\beta_{\mathrm{h}}$ also deserves further attention. Further research, including investigations into the causes of mode frequencies, and research into nonlinear phase, etc., will be presented in future work. We eagerly anticipate any future study providing additional insights into these subjects.

***Table 2 :*** **Effects of various terms in EP contributions on the IKM.** (For the parameters used in this study only.)

| | | Nearly Isotropic EPs | Trapped EPs | Passing EPs |
|---|---|---|---|---|
| **$\delta f$ contribution $(\nabla\cdot\delta \mathrm{P_h})_\perp^{(\delta f)}$** | | **Destabilizing** | **Destabilizing** | **Destabilizing** |
| | Adiabatic contribution $(\nabla\cdot\delta \mathrm{P_h})_\perp^{(\delta f,\mathrm{adi})}$ | **Destabilizing** | **Destabilizing** | Weakly **Stabilizing** |
| | Non-adiabatic contribution | Weakly **Stabilizing** | **Stabilizing** | **Destabilizing** |
| **$\delta B$ contribution: perp-terms $\delta[(\nabla\cdot \mathrm{P_h})_\perp]^{(\delta B,\mathrm{perp})}$** | | **Stabilizing** | **Stabilizing** | **Stabilizing** |
| **$\delta B$ contribution: cur-terms $\delta[(\nabla\cdot \mathrm{P_h})_\perp]^{(\delta B,\mathrm{cur})}$** | | Negligible | **Destabilizing** | **Stabilizing** |
| **Parallel components subtracted in $\delta f$ contribution** $(\nabla\cdot\delta \mathrm{P_h})_\parallel^{(\delta f)}$ | | **Stabilizing** | Weakly **Stabilizing** | **Strongly stabilizing** |
| **Total contributions of EPs in Scheme B** | | **Stabilizing** | **Stabilizing** | Weakly **Destabilizing** |

## Acknowledgements

This work is supported by the National MCF Energy R&D Program No. 2022YFE03100000, 2019YFE03030004, the National Natural Science Foundation of China No. 11835010, and US Department of Energy grants DE-AC02-09CH11466.

## ORCID iDs

Hanxiao Zhang https://orcid.org/0009-0007-1179-0350

Haowei Zhang https://orcid.org/0000-0003-1342-5756

Zhiwei Ma https://orcid.org/0000-0001-6199-9389

Chang Liu https://orcid.org/0000-0002-6747-955X

## Appendix A: Proof of Equivalence for Two Coupling of Complete Forms

When considering both $\delta f$ contribution and the $\delta \boldsymbol{B}$ contribution (as in Scheme B), the proof of equivalence between pressure-coupling and current-coupling schemes is as follows:

For the current-coupling scheme, the momentum equation subtracts the following terms related to EPs:

$$\begin{aligned}
&\delta \boldsymbol{J}_{\mathrm{h}} \times \boldsymbol{B} + \boldsymbol{J}_{\mathrm{h}0} \times \delta \boldsymbol{B} \\
&= [(\delta P_{\mathrm{h}\parallel} - \delta P_{\mathrm{h}\perp}) \nabla \times \boldsymbol{b} + \boldsymbol{b} \times \nabla \delta P_{\mathrm{h}\perp}] \times \boldsymbol{b} \\
&\quad + q_{\mathrm{h}} \left(\widetilde{N V}_{\mathrm{h}\parallel}\right)_0 (\delta \boldsymbol{b} \times \boldsymbol{B} + \boldsymbol{b}_0 \times \delta \boldsymbol{B}) \\
&\quad + \left[\left(\delta \boldsymbol{b} - \frac{\delta B}{B_0} \boldsymbol{b}_0\right) \times \nabla P_{\mathrm{h}\perp 0}\right] \times \boldsymbol{b} + (\boldsymbol{b}_0 \times \nabla P_{\mathrm{h}\perp 0}) \times \frac{\delta \boldsymbol{B}}{B_0} \\
&\quad + (P_{\mathrm{h}\parallel 0} - P_{\mathrm{h}\perp 0}) \left[\left(\nabla \times \delta \boldsymbol{b} - \frac{\delta B}{B_0} \nabla \times \boldsymbol{b}_0\right) \times \boldsymbol{b} + (\nabla \times \boldsymbol{b}_0) \times \frac{\delta \boldsymbol{B}}{B_0}\right] \\
&= [(\delta P_{\mathrm{h}\parallel} - \delta P_{\mathrm{h}\perp}) \nabla \times \boldsymbol{b} + \boldsymbol{b} \times \nabla \delta P_{\mathrm{h}\perp}] \times \boldsymbol{b} \\
&\quad + (\boldsymbol{b} \times \nabla P_{\mathrm{h}\perp 0}) \times \boldsymbol{b} - (\boldsymbol{b}_0 \times \nabla P_{\mathrm{h}\perp 0}) \times \boldsymbol{b}_0 \\
&\quad + (P_{\mathrm{h}\parallel 0} - P_{\mathrm{h}\perp 0}) [(\nabla \times \boldsymbol{b}) \times \boldsymbol{b} - (\nabla \times \boldsymbol{b}_0) \times \boldsymbol{b}_0] \\
&= (\nabla \delta P_{\mathrm{h}\perp})_\perp + (\delta P_{\mathrm{h}\parallel} - \delta P_{\mathrm{h}\perp})(\boldsymbol{b} \cdot \nabla) \boldsymbol{b} \\
&\quad + (\nabla P_{\mathrm{h}\perp 0})_\perp - (\nabla P_{\mathrm{h}\perp 0})_{\perp 0} + (P_{\mathrm{h}\parallel 0} - P_{\mathrm{h}\perp 0}) [(\boldsymbol{b} \cdot \nabla) \boldsymbol{b} - (\boldsymbol{b}_0 \cdot \nabla) \boldsymbol{b}_0]
\end{aligned} \tag{41}$$

In the current-coupling scheme, all terms involving $\left(\widetilde{N V}_{\mathrm{h}\parallel}\right)_0$ are automatically canceled out.

For the pressure-coupling scheme, the terms subtracted are as follows:

$$\begin{aligned}
&\delta[(\nabla \cdot \mathbf{P}_{\mathrm{h}})_\perp] \\
&= (\nabla \cdot \delta \mathbf{P}_{\mathrm{h}})_\perp + (\nabla \cdot \mathbf{P}_{\mathrm{h}0})_\perp - (\nabla \cdot \mathbf{P}_{\mathrm{h}0})_{\perp 0} \\
&= (\nabla \delta P_{\mathrm{h}\perp})_\perp + \{\nabla \cdot [(\delta P_{\mathrm{h}\parallel} - \delta P_{\mathrm{h}\perp}) \boldsymbol{b}\boldsymbol{b}]\}_\perp + (\nabla P_{\mathrm{h}\perp 0})_\perp - (\nabla P_{\mathrm{h}\perp 0})_{\perp 0} \\
&\quad + \{\nabla \cdot [(P_{\mathrm{h}\parallel 0} - P_{\mathrm{h}\perp 0}) \boldsymbol{b}\boldsymbol{b}]\}_\perp - \{\nabla \cdot [(P_{\mathrm{h}\parallel 0} - P_{\mathrm{h}\perp 0}) \boldsymbol{b}_0 \boldsymbol{b}_0]\}_{\perp 0} \\
&= (\nabla \delta P_{\mathrm{h}\perp})_\perp + (\delta P_{\mathrm{h}\parallel} - \delta P_{\mathrm{h}\perp})(\boldsymbol{b} \cdot \nabla) \boldsymbol{b} \\
&\quad + (\nabla P_{\mathrm{h}\perp 0})_\perp - (\nabla P_{\mathrm{h}\perp 0})_{\perp 0} + (P_{\mathrm{h}\parallel 0} - P_{\mathrm{h}\perp 0}) [(\boldsymbol{b} \cdot \nabla) \boldsymbol{b} - (\boldsymbol{b}_0 \cdot \nabla) \boldsymbol{b}_0]
\end{aligned} \tag{42}$$

In the pressure-coupling scheme, all terms containing $\nabla(P_{\mathrm{h}\parallel 0} - P_{\mathrm{h}\perp 0})$ are automatically canceled out as well.

Comparing Equations (41) and (42), their final forms are exactly the same, as shown in Equation (24).

## Appendix B: Toroidal Flow in Core Regions

In this $m/n = 1/1$ IKM case study, we observed that in certain situations, a significant $m/n = 1/1$ toroidal flow appears inside the rational surface of $q = 1$, as detected by $\delta V_{\phi}$. The emergence of this flow leads to an increase in mode growth rate and/or a decrease in frequency. Further investigation revealed that the occurrence of this flow depends not only on the choice of schemes within the model but also on whether EP pressure is included in the equilibrium pressure. Due to the complexity of the factors influencing the excitation of this toroidal flow, its mechanism remains unclear. We have documented the observations of this phenomenon in different cases in Figure 16 for comparison.

Simulations of nearly isotropic EPs using the M3D-C1-K code also observe the toroidal flow behavior that is completely consistent with the results from the CLT-K code. We present a comparison of partial results from CLT-K and M3D-C1-K codes in Figure 17.

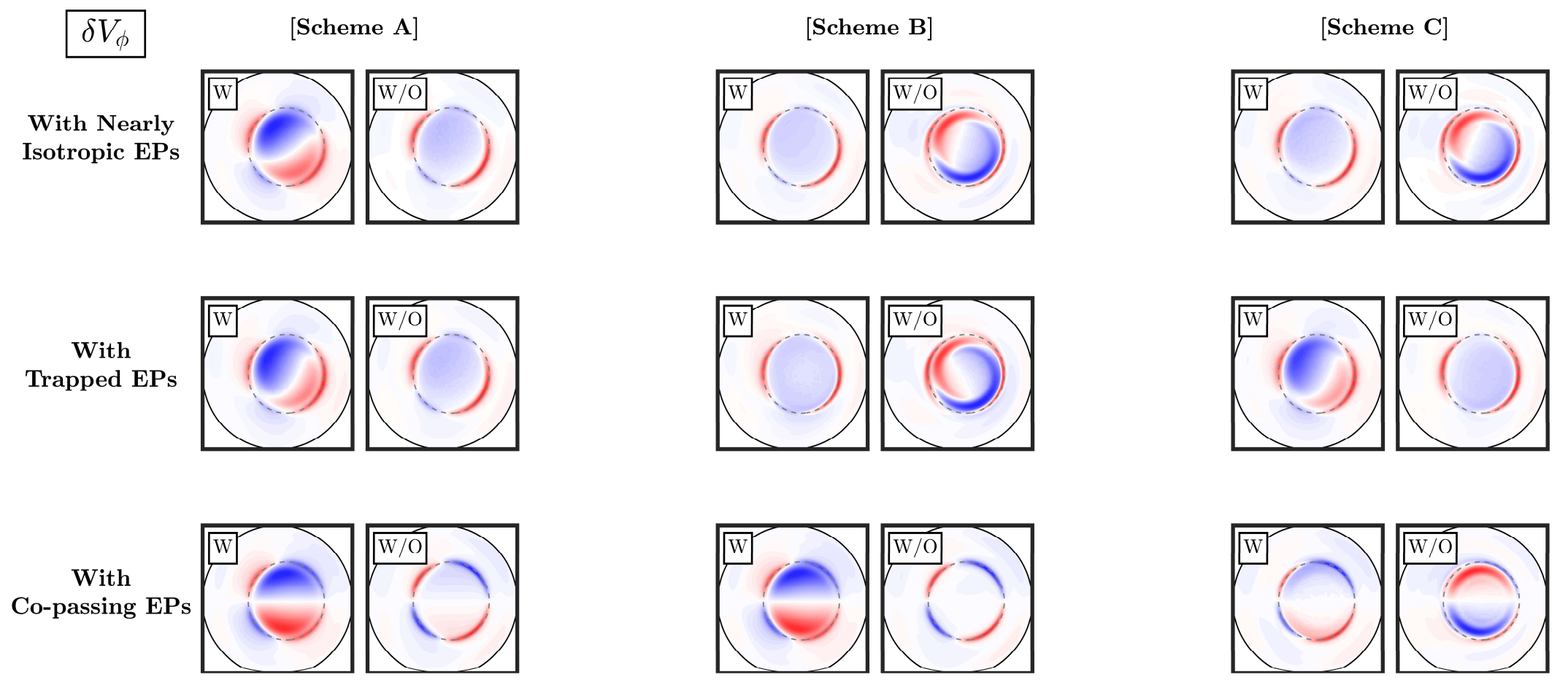

***Figure 16 :*** Structure of IKM toroidal fluid velocity $\delta V_{\phi}$ for different EP distributions ($\beta_{\mathrm{h}}/\beta_{\mathrm{total}} = 0.25$) and schemes. Labels "W" and "W/O" denote initial equilibrium with and without EP contributions, respectively.

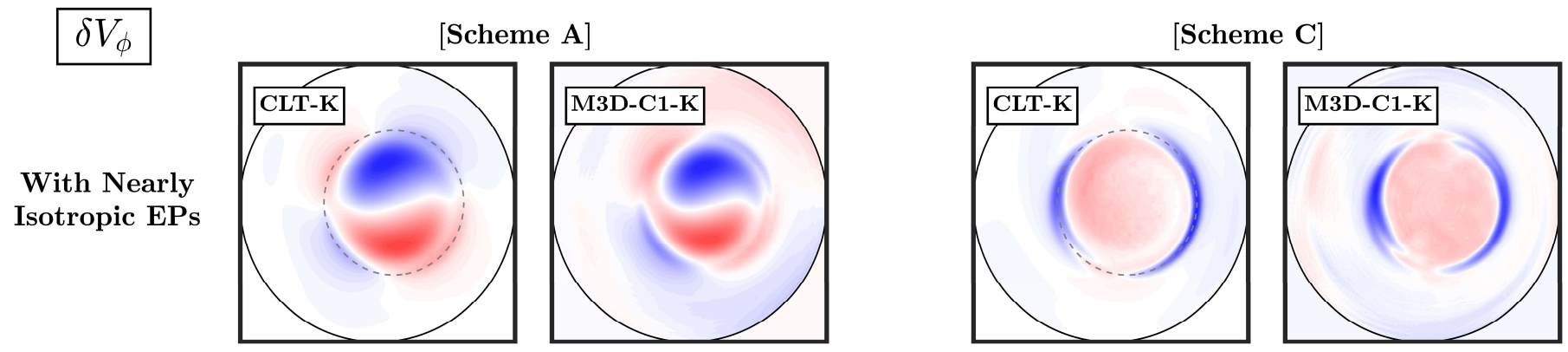

***Figure 17 :*** Structure of IKM toroidal fluid velocity $\delta V_{\phi}$ obtained from CLT-K and M3D-C1-K with nearly isotropic EPs ($\beta_{\mathrm{h}}/\beta_{\mathrm{total}} = 0.50$), using Scheme A and Scheme C respectively, with EP pressure contributions included in the initial equilibrium.